\documentclass[12pt]{article}
\usepackage[dvipsnames,svgnames]{xcolor}
\usepackage{amsfonts,amssymb,amsmath}
\usepackage{subfigure}
\usepackage{graphicx}
\usepackage{appendix}
\usepackage{enumerate}
\usepackage{tikz}
\usetikzlibrary{shapes,arrows,shadows,matrix}
\usetikzlibrary{shapes.gates.logic.US,trees,positioning}
\usepackage{fancyhdr}

            \newcommand{\av}{\alpha_v}
            \newcommand{\avi}{\alpha_v^{\text{i}}}
            \newcommand{\al}{\alpha_l}
            \newcommand{\li}{\lambda_i}
            \newcommand{\la}{\lambda}
            \newcommand{\lM}{\lambda_{max}}
            
            \newcommand{\lm}{\lambda_{min}}
            \newcommand{\lintM}{\lambda_{int}^{max}}

            \newcommand{\A}{\mathbb{A}}
            \newcommand{\pfp}{P_{HDD}}
            \newcommand{\pfppos}{P_{HDD}^{\text{pos}}}
            \newcommand{\Pds}{P_{HDF}}
            \newcommand{\Pexact}{P_{\text{exact}}}

            \newcommand{\V}{\mathbf{V}}
            \newcommand{\uv}{\mathbf{u}}

            \newcommand{\X}{\mathbb{X}}

            \newcommand{\Dt}{\Delta t}
            \newcommand{\ud}{\frac{1}{2}}
             \newcommand{\rv}{\rho_v}
             \newcommand{\rl}{\rho_l}
             \newcommand{\rk}{\rho_k}
            \newcommand{\Id}{\mathbb{I}}
            
            \newcommand{\setfm}{six equations two-fluid model }

\numberwithin{equation}{section}

\title{Phase appearance or disappearance \\in two-phase flows}

\author{Floraine Cordier$^{1,2,3}$, Pierre Degond$^{2,3}$, Anela Kumbaro$^{1}$ \\ 
  \small $\mbox{}$ \\ 
  \small $^{1}$ CEA-Saclay  \textsc{DEN, DM2S, SFME, LETR}  F-91191 Gif-sur-Yvette, France.\\
	\small floraine.cordier@cea.fr, anela.kumbaro@cea.fr\\
  \small  $^{2}$ Universit\'e de Toulouse; UPS, INSA, UT1, UTM ;\\ 
  \small Institut de Math\'ematiques de Toulouse ; F-31062 Toulouse, France. \\
  \small  $^{3}$ CNRS; Institut de Math\'ematiques de Toulouse UMR 5219 ; F-31062 Toulouse, France.\\
  \small pierre.degond@math.univ-toulouse.fr
}

\date{}

\begin{document}

\maketitle

\vspace{0.1 cm}

\paragraph{Abstract}
This paper is devoted to the treatment of specific numerical problems which appear when phase appearance or disappearance occurs in models of two-phase flows. Such models have crucial importance in many industrial areas such as nuclear power plant safety studies. In this paper, two outstanding problems are identified: first, the loss of hyperbolicity of the system when a phase appears or disappears and second, the lack of positivity of standard shock capturing schemes such as the Roe scheme. After an asymptotic study of the model, this paper proposes accurate and robust numerical methods adapted to the simulation of phase appearance or disappearance. Polynomial solvers are developed to avoid the use of eigenvectors which are needed in usual shock capturing schemes, and a method based on an adaptive numerical diffusion is designed to treat the positivity problems. An alternate method, based on the use of the hyperbolic tangent function instead of a polynomial, is also considered. Numerical results are presented which demonstrate the efficiency of the proposed solutions. 



\section{Introduction}
\label{sec:intro}

Multiphase flows can be found in a large variety of industrial or natural systems involving boiling or condensing fluids, reacting flows or aerosols. Such systems are, e.g., power plants, refrigerators, distillation units, gas or oil pipelines, pollutant separators, or clouds. The present work has been conducted in the context of nuclear power plant safety studies. In nuclear reactors, the appearance of vapor around the fuel rods interferes with the heat evacuation and can cause severe damages. To design and optimize the equipments in order to guarantee the highest possible safety level, numerical simulations of multiphase flows are intensively used. However, these simulations remain extremely delicate because of the complexity of the models and the possible huge discrepancy between the volume fraction of the various phases. 
For instance, within a subcooled liquid injected in a heated column, a transition from a single-phase liquid at the inlet to pure vapor at the outlet may take place. In such situations, numerical difficulties may be observed, like the loss of positivity of the mass fractions or internal energies. This is the case for instance with the CEA research code OVAP \cite{ovap} based on an implicit version of Roe's scheme. Therefore, a robust numerical scheme for two-phase flows must be able to treat all ranges of volume fractions.

In the literature, few works deal with the problem of phase appearance and disappearance explicitly. In most codes, this problem is treated using ad-hoc fixes. A first treatment has been developed in the "CATHARE" two-phase flow code \cite{cathare}. It relies on specific expressions of the interfacial mass and energy transfer terms which are designed such that the void fraction remains in an interval $[\alpha_{min}, \alpha_{max}]$. This treatment is combined with a numerical conditioning of the interfacial and wall friction source terms in order to provide a proper mechanical model for the coupling of the residual phases. A similar strategy is used in the "NEPTUNE" code \cite{guelfi2007neptune}. A second method is proposed in \cite{ausm} where an extension of the AUSM$+$ scheme (Advection Upwind Splitting Methods) to two-fluid models is developed. In \cite{ausm}, it is noticed that the AUSM$+$ numerical fluxes remain non-singular when a phase disappears as long as the involved Mach number and phase velocities remain bounded. Since it is assumed that the velocity  of the two phases should tend to each other at the transition, the velocities are therefore artificially tied to each other through a smooth function. A similar treatment is applied to the temperature. This treatment is applied when a phase has a volume fraction below $\alpha_{min}=10^{-4}$.

Therefore, the strategy developed in the literature is to treat these problems at the level of the underlying physics, by designing specific expressions for the interfacial closure terms. Without underestimating the role of the physics, we propose an alternative route. We explore the mathematical structure of the two-phase models in the limit of small volume fraction of one the phases.  This asymptotic approach is used to highlight the possible causes of the numerical breakdown and to design more robust methods. We restrict ourselves to models of two-phase flows but the methods could be extended to three of more phases. In the numerical investigations, we rely on a time-implicit version of the Roe scheme used in the "OVAP" code \cite{ovap}. We identify two essential difficulties, which are: (i) the loss of hyperbolicity of the two-fluid model when a phase appears or disappears, and (ii): the lack of positivity of the Roe scheme. Each of these difficulties will receive a specific treatment. 

To address the first difficulty, we propose the use of the so-called polynomial schemes \cite{P2}. This choice is motivated by the asymptotic analysis of the two-phase model in the limit of vanishing volume fraction of one of the phases. In this limit, the two phases almost decouple and the minority phase obeys a pressureless gas dynamics system \cite{PGD_Bouchut, 2004_NumericPGD_BouchutJin}. This system is not hyperbolic because the Jacobian of the flux matrix is not diagonalizable. This implies that two eigenvectors of the original two-phase model collapse in the limit of small volume fraction. Therefore, most shock-capturing schemes, which require a complete basis of eigenvectors, breakdown in this limit. To overcome this problem, schemes that do not require that eigenvectors form a complete basis are needed. There are many such schemes, such as Lax-Friedrichs, or central schemes \cite{KT}, but many of them are too diffusive for safety studies of nuclear power plants. The interesting feature with the polynomial schemes is that it is possible to tune the amount of numerical diffusion. Polynomial schemes have been used e.g. in \cite{ovap, thesemichael}. 

We will also consider an alternate method which uses the hyperbolic tangent function instead of a polynomial. It is as precise as the most precise polynomial method, and shows very good positivity properties without requiring any positivity treatment (see below). However, it is currently computationally too intensive for practical use. Nonetheless, improvements in the efficiency of the computation of the hyperbolic tangent function of a matrix could make this method potentially very competitive. 

The second difficulty, namely, the lack of positivity, is a critical issue in phase-transition problems. Indeed, they frequently appear in areas where the mass fraction of one of the phases is small. Then, small inaccuracies easily lead to negative mass fractions, especially with large time-steps. The simple fix consisting in replacing negative quantities by arbitrarily small positive values results in conservation losses and degraded robustness. For this reason, the development of positive schemes has been considered a major issue. In \cite{einfeldt1991godunov}, Einfeldt and al introduced the notion of "positively conservative" schemes where the density and internal energy remain positive. While the Godunov scheme is positively conservative, they show that no linearized Riemann solver, including the Roe scheme, is positively conservative. A more detailed bibliography about positive schemes can be found in section \ref{sec:pos}.

Several specific aspects make previously developed methods of difficult use for two-phase flow models, especially in the context of nuclear power plants safety. First, the models, such as the ones presented in the forthcoming sections, are complex non conservative systems. The analytical expressions of the eigenvalues and eigenvectors are not available. An analytical proof of the positivity of a scheme is therefore not possible. Additionally, such proofs strongly use the eigenstructure of the system. However, as explained above, this eigenstructure becomes singular when the volume fraction of one of the phases becomes small. As we will see in the review of section \ref{sec:pos}, many strategies leading to positive schemes are based on an increase of the numerical diffusion. But this additional diffusion is detrimental for the accuracy of the scheme, and accuracy is a critical issue for the targeted application. Another critical issue is efficiency and motivates the use of implicit schemes and large time-steps. In this context, schemes inducing positivity through a restriction on the CFL stability condition are not acceptable either. 

Our method does not guarantee positivity in all cases but, in practice, it solves most of the positivity problems while meeting the constraints listed above. The numerical treatment consists in an adaptive diffusion, which corrects positivity problems where they occur, locally in space and time. It is inspired from the works of Gallice \cite{gallice} and Romate \cite{romate1998approximate}, but the proposed strategy, which uses the framework of the polynomial schemes, with a specific choice of the polynomial, is, to our knowledge, original. 

The paper is organized as follows: section \ref{sec:asymptotic} develops the asymptotic study of the two-phase model when one of the phases disappears, showing that the model loses its hyperbolicity in this limit. Section \ref{sec:polyn} proposes the use of polynomial schemes in replacement of the Roe scheme to overcome the problem highlighted in section \ref{sec:asymptotic}. It provides a comparison between various choices for the polynomial and selects the most robust one. A method similar to polynomial solvers and using the hyperbolic tangent function is also detailed. Section \ref{sec:pos} addresses the positivity problem and proposes a new strategy to deal with it. Finally, numerical results are presented in section \ref{sec:numres}. A conclusion is given in section \ref{sec:conclu}. Auxiliary calculations are collected in appendix \ref{sec_appendix}.

\section{Two-phase flow models and phase appearance or disappearance}
\label{sec:asymptotic}

\subsection{The full two-phase model}
\label{sec:setfm}

This paper is concerned with two-phase flow models.  Detailed derivations and descriptions of two-phase flow models can be found in \cite{hestroni1982handbook, ishii75}.
In this section, we present the full two-phase model, including energy equations, which will be used in the numerical tests of section \ref{sec:numres}. Below, in section \ref{subsec_asymptotic}, an asymptotic analysis of the simpler, isentropic version of this system will be conducted.  

The unknown physical quantities are the volume fraction $\alpha_k \in [0,1]$, the density $\rho_k \geq 0$, the velocity $ u_k \in {\mathbb R}^d$, the energy $E_k \geq 0$, the enthalpy $h_k \geq 0$ of each of the phases indexed by $k$, where the subscript $k$ stands for $\ell$ for the liquid and $v$ for the vapor. They depend on position $x \in {\mathbb R}^d$ (where $d$ is the dimension), and time $t$. The common pressure of the two phases is denoted by $p$. Here, pressure equilibrium between the two phases is postulated. This hypothesis is known as the hydrostatic assumption. 
The model is written as follows, ignoring the viscous terms for simplicity:
\begin{eqnarray}
& & \hspace{-1cm}  
\partial_t ( \alpha_v \rho_v) + \nabla \cdot ( \alpha_v \rho_v u_v) = \Gamma \, , \label{res_mass_v2} \\
& & \hspace{-1cm}  
\partial_t (\alpha_\ell \rho_\ell) + \nabla \cdot (\alpha_\ell \rho_\ell u_\ell) = -\Gamma \, , \label{res_mass_l2} \\
& & \hspace{-1cm}  
\partial_t ( \alpha_v \rho_v u_v) + \nabla \cdot ( \alpha_v \rho_v u_v\otimes u_v) +  \alpha_v \nabla p +  D_{pi} \nabla  \alpha_v = \nonumber \\
& & \hspace{5cm}  
= \Gamma u^i +\alpha_v \rho_v f_{ext} + F^{iD}_v + F_w^v \, , \label{res_mom_v2} 
\end{eqnarray}

\begin{eqnarray}
& & \hspace{-1cm}  
\partial_t (\alpha_\ell \rho_\ell u_\ell) +\nabla \cdot (\alpha_\ell \rho_\ell u_\ell\otimes u_\ell) + \alpha_\ell \nabla p +  D_{pi} \nabla \alpha_\ell =  \nonumber \\
& & \hspace{5cm}  
= -\Gamma u^i +\alpha_\ell \rho_\ell f_{ext} + F^{iD}_\ell + F_w^l \, , \label{res_mom_l2} \\
& & \hspace{-1cm}
 \partial_t ( \alpha_v \rho_v E_v) + p \partial_t \alpha_v + \nabla \cdot ( \alpha_v \rho_v u_v (E_v + \frac{p}{\rho_v})) =  \nonumber \\
& & \hspace{1cm}
= \Gamma (\ud u_v^2 + h_v^i) + \alpha_v \rho_v f_{ext}\cdot u_v + Q_v^w + F^{iD}_v \cdot u^i \, , \label{res_en_v2} \\
& & \hspace{-1cm} 
\partial_t ( \alpha_\ell \rho_\ell E_\ell) + p \partial_t \alpha_\ell + \nabla \cdot ( \alpha_\ell \rho_\ell  u_\ell (E_\ell + \frac{p}{\rho_\ell})) =  \nonumber \\ 
& & \hspace{1cm}
= -\Gamma (\ud u_\ell^2 + h_\ell^i) +\alpha_\ell \rho_\ell f_{ext}\cdot u_\ell+ Q_\ell^w + F^{iD}_\ell \cdot u^i\, , \label{res_en_l2} 
\end{eqnarray}
\begin{eqnarray}
& &  \hspace{-1cm}  
\alpha_v + \alpha_\ell = 1 \, , \label{res_alpha2}\\
& & \hspace{-1cm} 
\rho_v = \rho_v(p,h_v) \, , \qquad h_v = E_v - \frac{u_v^2}{2} + \frac{p}{\rho_v} \, , \label{res_rho_v2} \\
& & \hspace{-1cm} 
\rho_\ell = \rho_\ell(p,h_\ell) \, , \qquad h_\ell = E_\ell - \frac{u_\ell^2}{2} + \frac{p}{\rho_\ell} \, . \label{res_rho_l2} 
\end{eqnarray}
where $D_{pi}$ is the interfacial pressure default proposed by Bestion \cite{bestion1990physical} and given by:  
\begin{equation}
  D_p^i =  \delta \, \alpha_v  \alpha_\ell  \,  \tilde \rho  \, |u_r|^2 .
  \label{term:bestion}
\end{equation}
The average density $\tilde \rho$ and the relative velocity $u_r$ are defined by
$$ \tilde \rho = \frac{ \rho_v  \rho_\ell}{\alpha_v \rho_\ell + \alpha_\ell \rho_v}, \qquad u_r = u_v-u_\ell, $$
and $\delta$ is an ad-hoc coefficient. $\rho_v(p,h_v)$ and $\rho_\ell(p,h_\ell)$ are the vapor and liquid equations-of-state. In the isentropic case (i.e. $\rho_v= \rho_v(p)$ and $\rho_\ell =\rho_\ell(p)$ only, see section \ref{subsec_asymptotic}), expression (\ref{term:bestion}) guarantees that the system is hyperbolic provided that $\delta \geq 1$ \cite{toumi1999ars}. 

The source terms have complex physical interpretations and we refer to \cite{hestroni1982handbook, ishii75} for details. They will not be discussed here. Specifically, $\Gamma$ is the interfacial mass transfer term, $u^i$ is the interfacial velocity, $f_{ext}$ is an external force such as gravity, $F^{iD}_k$ is the drag force, $F_w^k$ is the wall friction for each phase, $h_k^i$ is the interfacial liquid or vapor enthalpy, $ Q_k^w$ is the wall heat transfer for each phase. These terms are left undefined at this level because they depend on the specific test case. For each of the test case of section \ref{sec:numres}, their precise expression will be given.  

To analyze what occurs when the volume fraction of one of the phases becomes small, this model is too complex. Therefore, in the analysis section below, we focus on the isentropic model in one space dimension.

\subsection{Asymptotic analysis of the isentropic two-phase model}
\label{subsec_asymptotic}

We investigate the behavior of the isentropic two-phase model when the volume fraction of one of the phases vanishes. For the sake of simplicity, we consider the one-dimensional model and exclude any source or viscous terms except for the interfacial pressure default which makes the system hyperbolic. The isentropic two-fluid model is:

\begin{eqnarray}
&  &  \partial_t ( \alpha_v \rho_v) + \partial_x ( \alpha_v \rho_v u_v) = 0 \, , \label{res_mass_v1} \\
&  &  \partial_t (\alpha_\ell \rho_\ell) + \partial_x (\alpha_\ell \rho_\ell u_\ell) = 0 \, , \label{res_mass_l1} \\
&  &  \partial_t ( \alpha_v \rho_v u_v) + \partial_x ( \alpha_v \rho_v u_v^2) +  \alpha_v \partial_x p +  \delta \, \alpha_v  \alpha_\ell  \,  \tilde \rho  \,  u_r^2  \, \, \partial_x  \alpha_v = 0 \, , \label{res_mom_v1} \\
&  &  \partial_t (\alpha_\ell \rho_\ell u_\ell) + \partial_x (\alpha_\ell \rho_\ell u_\ell^2) + \alpha_\ell \partial_x p +  \delta \, \alpha_v  \alpha_\ell  \,  \tilde \rho  \,  u_r^2 \,  D_p^i \, \partial_x \alpha_\ell = 0 \, , \label{res_mom_l1} \\
&  &  \rho_v = \rho_v(p) \, , \label{res_rho_v1} \\
&  &  \rho_\ell = \rho_\ell(p) \, , \label{res_rho_l1} \\
&  &   \alpha_v + \alpha_\ell = 1 \, , \label{res_alpha1}
\end{eqnarray}
This model is hyperbolic provided $\delta \geq 1$ \cite{toumi1999ars}. 

Let us now focus on the behaviour of the system when a phase disappears. We consider for instance that the vapor phase is disappearing. The vapor volume fraction $\av$ becomes close to zero. Therefore, it is legitimate to introduce a small parameter $\varepsilon \ll 1$ which measures the order of magnitude of $\alpha_v$ and to rescale $\alpha_v$ as follows: 
\begin{equation}
  \alpha_v = \varepsilon \bar \alpha_v. \label{rescale}
\end{equation}
After this rescaling, the system becomes: 
\begin{eqnarray}
& & \partial_t (\bar \alpha_v \rho_v) + \partial_x (\bar \alpha_v \rho_v u_v) = 0 \, , \label{res_mass_v} \\
& & \partial_t (\alpha_\ell \rho_\ell) + \partial_x (\alpha_\ell \rho_\ell u_\ell) = 0 \, , \label{res_mass_l} \\
& & \partial_t (\bar \alpha_v \rho_v u_v) + \partial_x (\bar \alpha_v \rho_v u_v^2) + \bar \alpha_v \partial_x p + \varepsilon \, \bar \alpha_v \alpha_\ell \tilde \rho  \, u_r^2  \,  \delta \, \partial_x \bar \alpha_v = 0 \, , \label{res_mom_v} \\
& & \partial_t (\alpha_\ell \rho_\ell u_\ell) + \partial_x (\alpha_\ell \rho_\ell u_\ell^2) + \alpha_\ell \partial_x p + \varepsilon  \, \bar \alpha_v \alpha_\ell \tilde \rho \,  u_r^2  \, \delta \, \partial_x \alpha_\ell = 0 \, , \label{res_mom_l} \\
& & \rho_v = \rho_v(p) \, , \label{res_rho_v} \\
& & \rho_\ell = \rho_\ell(p) \, , \label{res_rho_l} \\
& & \varepsilon \bar \alpha_v + \alpha_\ell = 1 \, . \label{res_alpha}
\end{eqnarray}
We now write the system obeyed by the formal limit $\varepsilon \to 0$: 
\begin{eqnarray}
& & \hspace{-3cm} \partial_t (\bar \alpha_v \rho_v) + \partial_x (\bar \alpha_v \rho_v u_v) = 0 \, , \label{lim_mass_v} \\
& & \hspace{-3cm} \partial_t \rho_\ell + \partial_x (\rho_\ell u_\ell) = 0 \, , \label{lim_mass_l} \\
& & \hspace{-3cm} \partial_t (\bar \alpha_v \rho_v u_v) + \partial_x (\bar \alpha_v \rho_v u_v^2) + \bar \alpha_v \partial_x p = 0 \, , \label{lim_mom_v} \\
& & \hspace{-3cm} \partial_t (\rho_\ell u_\ell) + \partial_x ( \rho_\ell u_\ell^2) + \partial_x p  = 0 \, , \label{lim_mom_l} \\
& & \hspace{-3cm} \rho_v = \rho_v(p) \, , \label{lim_rho_v} \\
& & \hspace{-3cm} \rho_\ell = \rho_\ell(p) \, . \label{lim_rho_l} 
\end{eqnarray}

Let us make a few comments on the structure of the limit system. First, we notice that the system composed of eqs. (\ref{lim_mass_l}), (\ref{lim_mom_l}) and (\ref{lim_rho_l}) is nothing but the isentropic Euler system for a single fluid consisting of the liquid phase. Indeed, the isentropic pressure of this fluid is given by the inverse function $p(\rho_\ell)$ of $\rho_\ell(p)$. The system for the liquid phase is thus completely decoupled from the vapor phase. 

Let us now turn towards the system consisting of eqs. (\ref{lim_mass_v}), (\ref{lim_mom_v}), (\ref{lim_rho_v}) which determines the vapor variables. Since the pressure $p$ is entirely determined by the liquid phase, the pressure term $\bar \alpha_v \partial_x p$ in (\ref{lim_mom_v}) is a zero-th order term in $\alpha_v$, multiplied by a known coefficient $\partial_x p$. Therefore, the system for the liquid variables can be written 
\begin{align}
   & \partial_t (\bar \alpha_v \rho_v) + \partial_x (\bar \alpha_v \rho_v u_v) =0 \, , \label{eq_vap_rho}\\
   & \partial_t (\bar \alpha_v \rho_v u_v) + \partial_x (\bar \alpha_v \rho_v u_v^2) = S_v\, , \label{eq_vap_u}
\end{align}
where $S_v$ contains only zero-th order terms. The hyperbolicity of the model is determined by the left-hand sides of (\ref{eq_vap_rho}), (\ref{eq_vap_u}). The corresponding system is a pressureless gas dynamics system for the variable $U = (\bar \alpha_v \rho_v, \bar \alpha_v \rho_v u_v )$. The pressureless gas dynamics system is  not hyperbolic. If we write this system $ \partial_t U + \partial_x f(U)=S $, with $f(U) = (\bar \alpha_v \rho_v u_v, \bar \alpha_v \rho_v u_v^2 )$ and $S = (0,S_v)$, the Jacobian matrix $\frac{\partial f}{\partial U}$ does not have a complete basis of eigenvectors. More precisely, $u_v$ is an eigenvalue of multiplicity $2$ but the associated eigenspace is of dimension $1$. The matrix $\frac{\partial f}{\partial U}$ can be written in the form of a Jordan block of size $2$. We refer to \cite{PGD_Bouchut, 2004_NumericPGD_BouchutJin} for a detailed analysis of the pressureless gas dynamics equations. 

Now, we consider the scaled system (\ref{res_mass_v}), (\ref{res_alpha}). It is a strictly hyperbolic $4 \times 4$ system \cite{toumi1999ars}. Consequently, it has a complete basis of $4$ eigenvectors. In the limit $\varepsilon \to 0$, two of these eigenvectors converge towards corresponding eigenvectors of the isentropic Euler system for the liquid phase. The other two eigenvectors become parallel to each other and parallel to the unique eigenvector of the vapor phase pressureless gas system. Appendix \ref{append:eigenvectors} confirms this deduction: using the first-order approximation of the eigenvectors given in \cite{toumi96} for the perfect gas equation-of-state, we show that the eigenvectors corresponding to the void fraction and pressure waves become parallel to each other when $\av \to 0$. 

To summarize, this analysis shows that, when a phase disappears, some eigenvectors collapse and become parallel. We will see that this phenomenon can raise some issues for the numerical scheme.

\subsection{Roe scheme and phase appearance / disappearance}
\label{subsec_Roe}

Roe's approximate Riemann solver \cite{roe1981approximate, toro2009riemann} is one of the most powerful and widely used schemes to solve hyperbolic systems of conservation laws. However, the two-fluid model has non-conservative terms. Toumi and Kumbaro \cite{toumi1999ars} have proposed a generalization of the Roe linearization to non-conservative systems.
The non-conservative two-phase system can be written in the quasi-linear form: 
\begin{equation}
  \frac{\partial \V}{\partial t} + \A(\V) \frac{\partial \V}{\partial x} = 0.
\end{equation}
In the finite volume framework, the generalized Roe scheme can be written as:
\begin{eqnarray}
  \frac{\V^{n+1}_i - \V^{n}_i}{\Delta t} \quad+\quad \frac{1}{\Delta x} \left( \Phi^{-}(\V_i,\V_{i+1}) + \Phi^+(\V_{i-1},\V_{i}) \right)  = 0,
  \label{ncons_scheme}
\end{eqnarray}
\noindent with
\begin{equation}
  \Phi^{\pm}(\V_i,\V_{i+1}) =  \A^{\pm}(\tilde\V_{i+\ud}) (\V_{i+1} - \V_i).
\end{equation}
The Roe matrix is the Jacobian matrix $\A$ of the system taken in an appropriate linearization state $\tilde\V_{i+\ud}$. For a non-conservative system, the linearization state is chosen so that the shock waves at the interface between cells $i$ and $i+1$ remain those of an equivalent conservative system \cite{toumi1999ars}. The positive and negative Roe matrices are defined by:
\begin{equation}
  \A^{\pm}(\tilde\V_{i+\ud}) = \frac{\A(\tilde\V_{i+\ud}) \pm |\A(\tilde\V_{i+\ud})|}{2} \, .
  \label{eq_Apm}
\end{equation}
where $|\A(\tilde\V_{i+\ud})|$ is the absolute value of $\A(\tilde\V_{i+\ud})$. 
In the OVAP code, the second term in (\ref{ncons_scheme}) is evaluated implicitly and the resulting nonlinear system for $V_i^{n+1}$ is solved by Newton's iterations \cite{ovap}. 

The computation of the absolute value of $|\A(\tilde\V_{i+\ud})|$ is performed as follows. Let $\A$ be a diagonalizable matrix. We write 
\begin{equation}
\A= {\mathcal R} \,  \text{diag}(\la_1 , \ldots, \la_N) \,  {\mathcal R}^{-1}, 
\label{eq_A_decomp}
\end{equation}
where the $\lambda_k$ 's are the eigenvalues of $\A$, $\text{diag}(\la_1,\ldots,\la_N)$ is the diagonal matrix whose diagonal coefficients are the $\la_k$ 's, and  ${\mathcal R}$ is the matrix whose columns are the eigenvectors of $\A$. Then, $|\A|$ is given by
\begin{equation}
|\A| = {\mathcal R} \, \text{diag}(|\la_1| , \ldots , |\la_N|) \, {\mathcal R}^{-1}. 
\label{eq_matrix_abs}
\end{equation}

Formula (\ref{eq_matrix_abs}) for the matrix absolute value is valid as long as $\A$ is diagonalizable. However, if the system loses its hyperbolicity, the eigenvectors of the Jacobian matrix $\A$ do not form a complete basis anymore, the matrix ${\mathcal R}$ becomes singular and strictly speaking, $|\A|$ is no more defined. We have seen that the limit system (\ref{res_mass_v1})-(\ref{res_alpha1}) is not hyperbolic for $\av \to 0$ for the precise reason that the eigenvectors do not form a complete basis any longer. In practice, during a computation, numerical problems begin to appear with the Roe scheme for $\av \in [10^{-2}, 10^{-4}]$, depending on the considered case. These problems are caused by some of the eigenvectors becoming almost parallel when the volume fraction decreases. The matrix ${\mathcal R}$ becomes highly ill-conditioned. The numerical accuracy of the eigenvector decomposition is then strongly affected. Therefore, the use of the Roe scheme based on an eigenvector decomposition of the Roe matrix must be avoided when phases appear or disappear. We will see that $|\A|$ can be computed with different methods which do not require the use of the eigenvector decomposition of $\A$. With this aim, we recall a certain number of results stemming from functional calculus

Let $\A$ be a diagonalizable matrix and denote by $\mbox{Sp}(\A) = \{ \lambda_1, \ldots , \lambda_N \}$ the spectrum of $\A$. Let $\Phi$ be a continuous function defined on an open interval ${\mathcal I}$ containing $\mbox{Sp} (\A)$. The matrix $\Phi (\A)$ is defined by 
$$ \Phi(\A) = {\mathcal R} \, \text{diag}(\Phi(\la_1),\cdots,\Phi(\la_N)) \,   {\mathcal R}^{-1}, $$
with ${\mathcal R}$ defined by (\ref{eq_A_decomp}). We note that $\Phi (\A)$ only depends on the values of $\Phi$ on $\mbox{Sp}(\A)$. Additionally, if $\Phi_n$ is a sequence of function such that $(\Phi_n(\lambda_1), \ldots, \Phi_n(\lambda_N)) \to (\Phi(\lambda_1), \ldots, \Phi(\lambda_N))$ in ${\mathbb R}^N$, then $ \Phi_n(\A) \to \Phi(\A)$ in any matrix norm. Of course, this is the case if $\|\Phi_n - \Phi\|_\infty \to 0$. Here, $\| \Phi \|_\infty$ denotes the uniform norm in the space $C^0(\bar {\mathcal I})$ of continuous functions on the closure $\bar {\mathcal I}$ of ${\mathcal I}$. 

Consequently, if a function $\Phi(x)$ approximates the absolute function $|x|$, the resulting $\Phi(\A)$ approximates $|\A|$ to the same order. Thus, we are looking for approximation functions $\Phi$ which allow the computation of $\Phi(\A)$ without requiring the eigenvector decomposition (\ref{eq_A_decomp}). 
This can be achieved by taking $\Phi$ as a polynomial $P$ such that $P(\lambda_i) \approx |\lambda_i|$, for all $i=1,\ldots,N$. Indeed, $P(\A)$ can be simply calculated by taking successive powers $\A^k$ of $\A$ and does not require the eigenvector decomposition. This gives rise to the so-called polynomial schemes \cite{P2}. Then, we will also consider an alternative, consisting in using the hyperbolic tangent function, which can be computed by solving a matrix ordinary differential equation. In all these cases, $\Phi(\A)$ will still be defined even when $\A$ ceases to be diagonalizable and the scheme will not breakdown at phase appearance or disappearance.

\subsection{Eigenvalues of the full two-phase model}
\label{subsec_eigen_full}

Although the full two-phase model of section \ref{sec:setfm} is not as simple as the isentropic model of section \ref {subsec_asymptotic}, some information about the eigenvalues and eigenvectors of the system can be obtained. Because of the complexity of this model, no analytical expression of the eigenvalues is available. However, approximations given in \cite{KTC, TK96} enable us to discuss the behavior of the eigenvalues when a phase appears or disappears. The detailed computation is given in appendix \ref{append:eigenvalues}. 

Since the hyperbolicity of the model is only determined by the left-hand sides of eqs (\ref{res_mass_v2})- (\ref{res_en_l2}), the precise knowledge of the source term is again unnecessary. The system is posed in dimension $d$. So, there are $4+2d$ eigenvalues of the Jacobian matrix. In general, there are two fast eigenvalues which are of the order of $u_{v} \pm c$, where $c$ is a characteristic sound velocity of the two-phase mixture, two eigenvalues of the order of $u_{\ell}$ called the void eigenvalues, and  two trivial eigenvalues, each of multiplicity $d$, respectively equal to the vapor and liquid velocities $u_{v}$ and $u_{\ell}$. Note that the void eigenvalues can be complex if the interfacial closure terms are not carefully chosen (see \cite{thesemichael}). The fastest eigenvalues $u_{v } \pm c$ are always real and remain distinct from the other eigenvalues. We will denote them by $\lM$ for the largest and $\lm$ for the smallest. All the other eigenvalues, that we will call the "intermediate eigenvalues" and collectively denote by $ \la_k^{\text{intermediate}}$ have the same orders of magnitude as long as the two-phase flow stays subsonic.  In the example of a boiling channel which is relevant for our applications (see  section \ref{sec:numres}), the ratios between the orders of magnitude of the fastest eigenvalues and the intermediate eigenvalues are the following:
\begin{equation}
  \frac{|\la_k^{\text{intermediate}}|}{|\lM|}\approx  \frac{|\la_k^{\text{intermediate}}|}{|\lm|}\approx 10^{-4}
  \label{eq:order_eigen}
\end{equation}

Suppose now that the vapor phase disappears : $\av\to 0$. 
Then, the fast eigenvalues tend towards $u_{\ell} \pm a_m$, where the expression of $a_m$ is given in appendix \ref{append:eigenvalues}. They remain distinct and the associated eigenvectors do not collapse. However, the void eigenvalues become of the order of magnitude of $u_{v}$ and the corresponding eigenvectors collapse. Qualitatively, the same phenomenon as in the isentropic  case occurs: two eigenvectors become parallel in the limit  $\av\to 0$ and the eigenvectors do not form a complete basis any longer. The matrix ${\mathcal R}$ formed by the eigenvectors becomes ill-conditioned. The computation of the Roe matrix becomes highly inaccurate.

\section{Numerical schemes based on polynomial or hyperbolic tangent evaluations of $\A$}
\label{sec:polyn}

As already announced, polynomial schemes avoid the use of the eigenvector decomposition of the Roe matrix $\A$ to compute $|\A|$. In this section, we give a presentation of polynomial schemes and provide a selection of high-degree polynomials which are well-suited to multi-phase flow calculations in the situation of phase appearance or disappearance. Polynomial schemes have been introduced in \cite{P2} and used in \cite{ovap, thesemichael}. We also present an alternative, based on the evaluation of $\A$ using the hyperbolic tangent function. This method is, to the best of our knowledge, new.

\subsection[Computation of |A| with a polynomial]{Computation of $|\A|$ with a polynomial}
\label{approx_polyn}

We recall the approach sketched at the end of section \ref{subsec_Roe}. It relies on the approximation of $|\A|$ by a polynomial $P$ such that $P(\A) \approx |\A|$. Indeed, the matrix polynomial 
\begin{equation}
P(\A) = \sum_{k=0}^n a_k \A^k,
\label{eq_P(A)_0}
\end{equation} 
of a diagonalizable matrix $\A$ can be alternately computed, using the eigenvector decomposition (\ref{eq_A_decomp}), by: 
\begin{equation}  
P(\A) = {\mathcal R}  \, \text{diag}(P(\la_1),\ldots, P(\la_N)) \,  {\mathcal R}^{-1}. 
\label{eq_P(A)}
\end{equation}
Therefore, if $P$ satisfies 
\begin{equation}
  P(\li)=|\li| , \qquad \forall i \in [1 \ldots N] , 
  \label{cond_f}
\end{equation}
i.e. if it interpolates the absolute value function at all the eigenvalues of $\A$, then 
\begin{equation}
P(\A)=|\A|. 
\label{eq_P(A)_2}
\end{equation}
Therefore, there are two ways of computing $|\A|$: either by formula (\ref{eq_matrix_abs}), or by (\ref{eq_P(A)_0}) with a polynomial $P$ satisfying (\ref{cond_f}). However, the advantage of formula (\ref{eq_P(A)_0}) over (\ref{eq_matrix_abs}) is that it does not use the eigenvector matrix ${\mathcal R}$ and is consequently faster. Additionally, the computation of $P(A)$ does not breakdown if the matrix is not diagonalizable, while that of $|\A|$ does. In fact, $|\A|$ is no more defined in this case while $P(A)$ stays defined. Therefore, the polynomial formula for $|\A|$ is better suited to the case where the eigenvector decomposition of $\A$ breaks down and the matrix ${\mathcal R}$ becomes ill-conditioned. In view of the discussion of section \ref{subsec_Roe}, polynomial schemes appear as methods of choice for situations of phase appearance or disappearance

In practice, the selection of the polynomial is crucial. Indeed, it may be useless to verify (\ref{cond_f}) exactly, i.e. for all the eigenvalues. It may increase computational costs to no avail and may be detrimental to the stability of the scheme. If (\ref{cond_f}) is not satisfied exactly, then $P(\A) \approx |\A|$ instead of satisfying (\ref{eq_P(A)_2}) exactly. The selection of $P$ becomes a compromise between accuracy on the one hand, and stability and computational efficiency on the other hand. We discuss these issues below.

\begin{figure}[h!]
  \centering
  \begin{tikzpicture}[>=stealth,scale=0.6] 
    \draw[->] (-3,0) -- (3,0) node[below,font=\tiny] {$x$};
    \draw[->] (0,-1) -- (0,3) node[left,font=\tiny] {$y$};
    \draw[Purple,dashed] (-3,3) -- (0,0) -- (3,3) node[above,font=\tiny] {$y=|x|$};
    \draw (2,2) -- (2,0) node[below,font=\tiny] {$a_{\max}$};
    \filldraw (2,2) circle (1pt);
    \draw[color=black] (-2,2) -- (2,2) node[right,font=\tiny] {};
    \fill[color=yellow!50!white,fill opacity=0.5] (-2,2) -- (0,0) -- (2,2) -- cycle;
  \end{tikzpicture}
  \caption{Stability condition. The graph of the polynomial in the interval $[\lambda_{min},\lambda_{max}]$ must be contained in the coloured area in order to ensure the stability of the explicit scheme.}
  \label{graph:stability}
\end{figure}
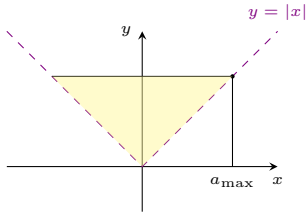

For explicit schemes, Degond and al \cite{P2} have shown a sufficient $L^2$ stability condition for polynomial schemes, under the CFL condition. Let $\lambda_{\min}$ and $\lambda_{\max}$ be the smallest and largest eigenvalues of the Roe matrix $\A$, and $a_{\max} = \max\{|\lambda_{\min}|,|\lambda_{\max}|\}$. Then, the stability criterion reads
\begin{equation}
|x| \leq P(x) \leq a_{\max}, \qquad \forall x \in [\lambda_{min},\lambda_{max}]. 
\label{eq_stabi_poly}
\end{equation}
This condition is represented graphically in Fig.\ref{graph:stability}. The graph of the polynomial in the interval $[\lambda_{min},\lambda_{max}]$ must be contained in the coloured area of the figure. 
 
\medskip

\begin{figure}
  \centering
  \subfigure[$\Pexact$ \label{fig:RFb}]{\includegraphics[width=0.43\textwidth,trim=30 50 80 80,clip]{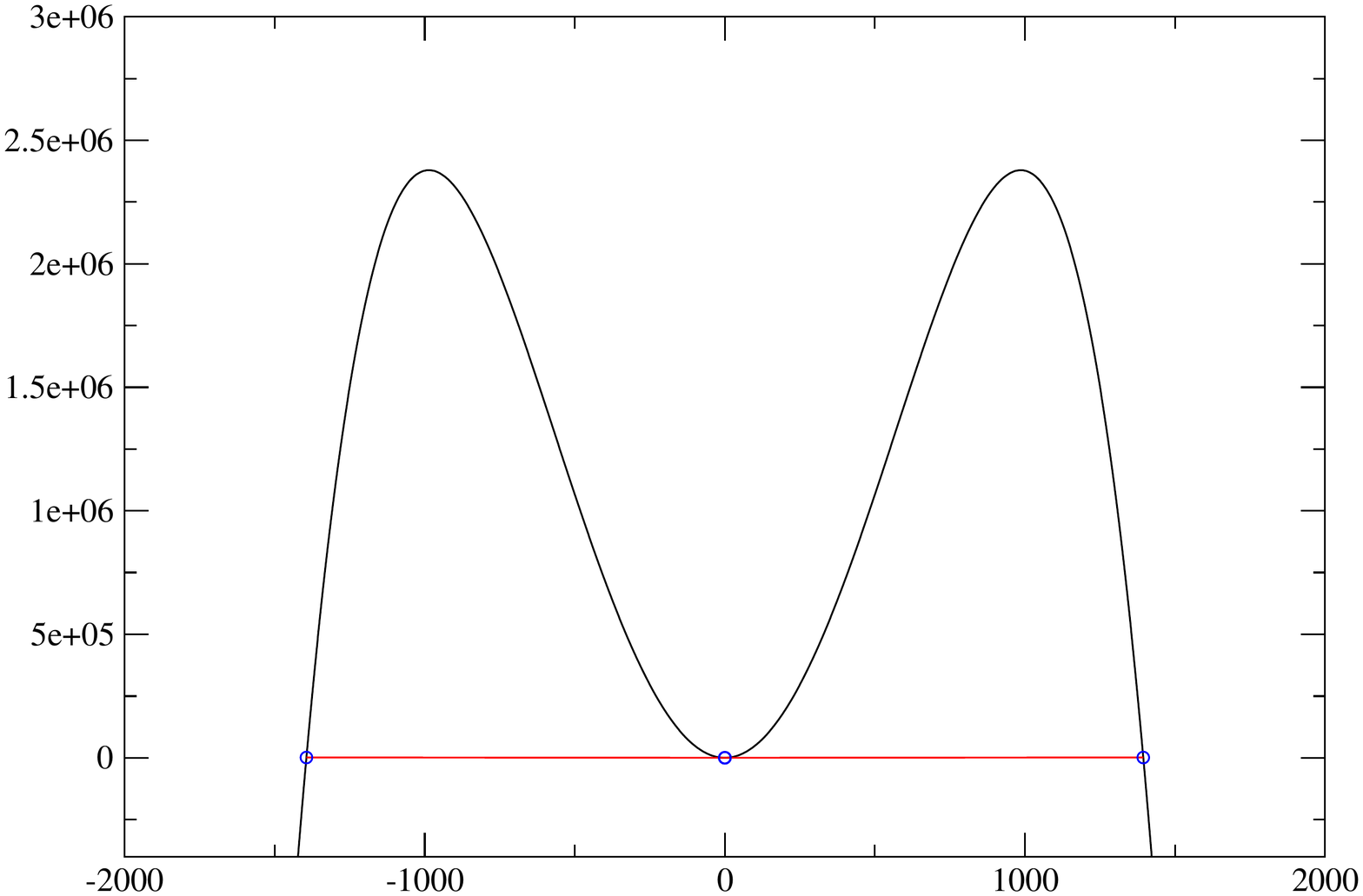}}
  \subfigure[Zoom on small eigenvalues\label{fig:RFc}]{\includegraphics[width=0.43\textwidth,trim=30 50 80 80,clip]{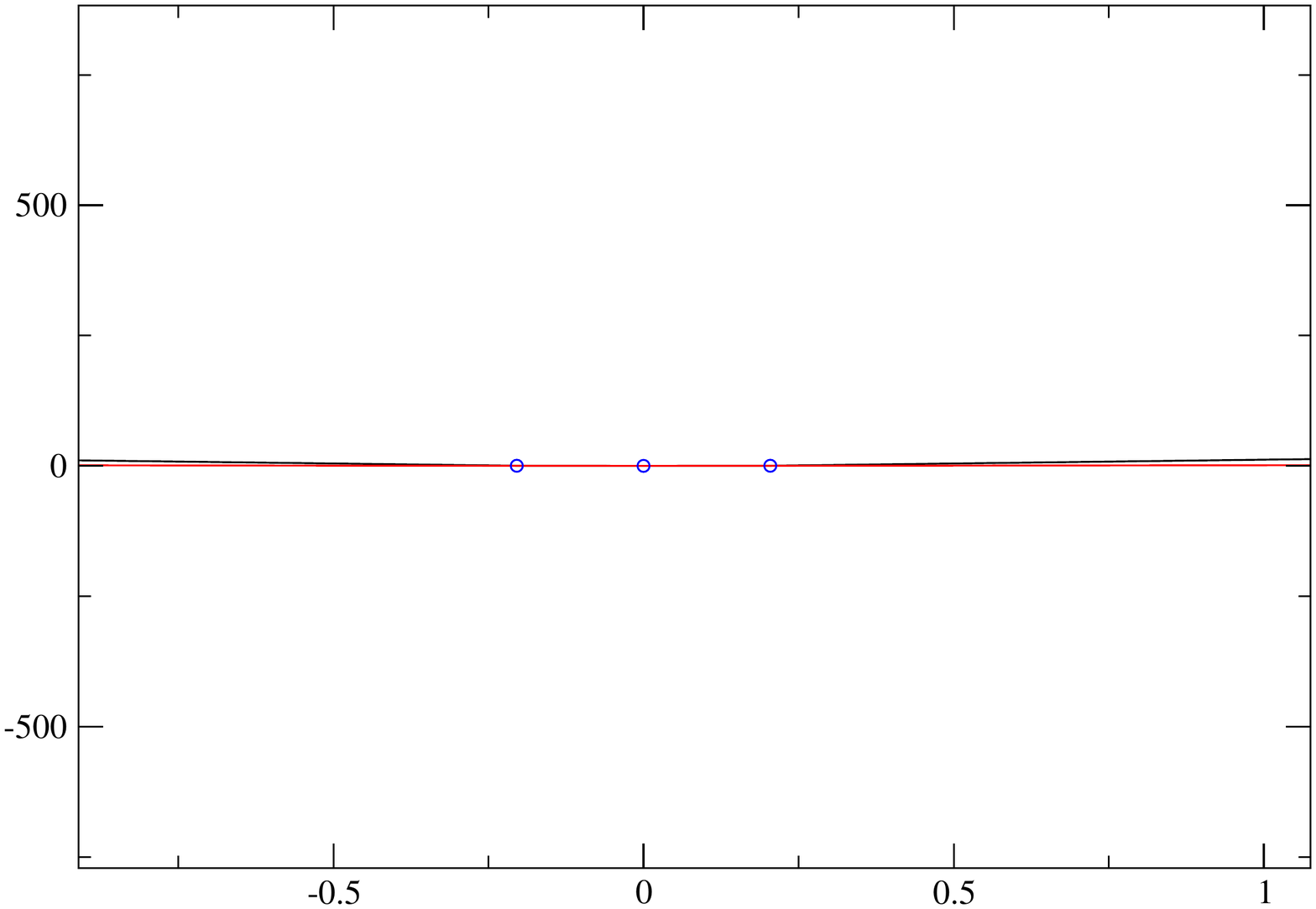}}
  \caption{Exact interpolating polynomial $\Pexact$. The black line is the polynomial, the red line is the absolute value function, and the blue spots are the eigenvalues. Left, the graph of the polynomial in the full range of interest $[\lambda_{min}, \lambda_{max}]$. Right: a blow-up of the graph in the region of eigenvalues of the smallest absolute values. }
  \label{fig:p_exact}
\end{figure}

Condition (\ref{eq_stabi_poly}) ensures the stability of the scheme. Accuracy requires that large oscillations of the polynomials near the eigenvalues should be avoided. Indeed, if the derivative of the polynomial about one of the eigenvalues is large, round-off errors may be amplified. A small difference between the true eigenvalue $\la$ and the computed one $\tilde \la$  may cause a huge discrepancy between $|\la|$ and $P(\tilde \la)$, thus creating numerical inaccuracies. In  \cite{ovap, thesemichael}, the Lagrange interpolation of the $|\lambda_i|$ 's is used in the Newton basis. This interpolating polynomial, further on referred to as '$\Pexact$'  verifies (\ref{cond_f}) exactly but has large oscillations. The resulting method is as good as the classical Roe scheme in standard situations. However, it breaks down at phase appearance or disappearance due to the large oscillations that are generated at the extremal eigenvalues (see fig. \ref{fig:p_exact}). These oscillations are caused by the intermediate eigenvalues which get very close one to each other (see section \ref{sec:asymptotic}). This induces ill-conditioning of the matrix involved in Newton's method of computation of the Lagrange interpolation polynomial and loss of accuracy. In what follows, we develop new approximating polynomial avoiding this difficulty.

\medskip

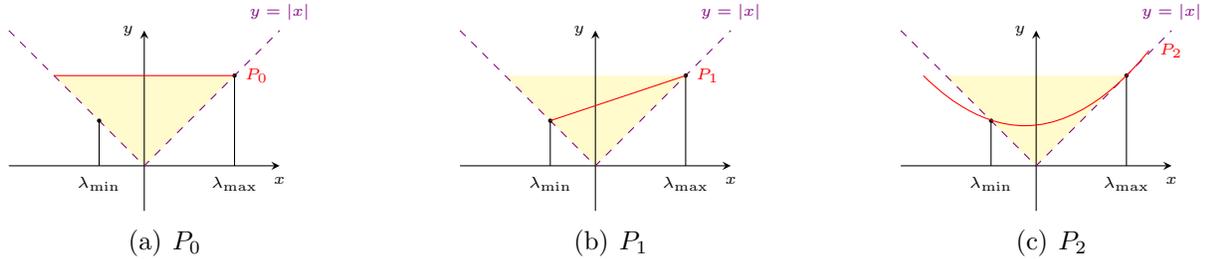
\begin{figure}[h!]
  \subfigure[$P_0$ ]{\begin{tikzpicture}[>=stealth,scale=0.6] 
      \fill[color=yellow!50!white,fill opacity=0.5] (-2,2) -- (0,0) -- (2,2) -- cycle;
      \draw[->] (-3,0) -- (3,0) node[below,font=\tiny] {$x$};
      \draw[->] (0,-1) -- (0,3) node[left,font=\tiny] {$y$};
      \draw[Purple,dashed] (-3,3) -- (0,0) -- (3,3) node[above,font=\tiny] {$y=|x|$};
      \draw (-1,1) -- (-1,0) node[below,font=\tiny] {$\lambda_\mathrm{min}$};
      \draw (2,2) -- (2,0) node[below,font=\tiny] {$\lambda_\mathrm{max}$};
      \filldraw (-1,1) circle (1pt);
      \filldraw (2,2) circle (1pt);
      \draw[color=red] (-2,2) -- (2,2) node[right,font=\tiny] {$P_0$};
  \end{tikzpicture} }
  \hspace{\stretch{1}}
  \subfigure[$P_1$ ]{\begin{tikzpicture}[>=stealth,scale=0.6] 
      \fill[color=yellow!50!white,fill opacity=0.5] (-2,2) -- (0,0) -- (2,2) -- cycle;
      \draw[->] (-3,0) -- (3,0) node[below,font=\tiny] {$x$};
      \draw[->] (0,-1) -- (0,3) node[left,font=\tiny] {$y$};
      \draw[Purple,dashed] (-3,3) -- (0,0) -- (3,3) node[above,font=\tiny] {$y=|x|$};
      \draw (-1,1) -- (-1,0) node[below,font=\tiny] {$\lambda_\mathrm{min}$};
      \draw (2,2) -- (2,0) node[below,font=\tiny] {$\lambda_\mathrm{max}$};
      \filldraw (-1,1) circle (1pt);
      \filldraw (2,2) circle (1pt);
      \draw[color=red] (-1,1) -- (2,2) node[right,font=\tiny] {$P_1$};
  \end{tikzpicture}}
  \hspace{\stretch{1}}
  \subfigure[$P_2$ ]{\begin{tikzpicture}[>=stealth,scale=0.6]
      \fill[color=yellow!50!white,fill opacity=0.5] (-2,2) -- (0,0) -- (2,2) -- cycle;
      \draw[->] (-3,0) -- (3,0) node[below,font=\tiny] {$x$};
      \draw[->] (0,-1) -- (0,3) node[left,font=\tiny] {$y$};
      \draw[Purple,dashed] (-3,3) -- (0,0) -- (3,3) node[above,font=\tiny] {$y=|x|$};
      \draw (-1,1) -- (-1,0) node[below,font=\tiny] {$\lambda_\mathrm{min}$};
      \draw (2,2) -- (2,0) node[below,font=\tiny] {$\lambda_\mathrm{max}$};
      \filldraw (-1,1) circle (1pt);
      \filldraw (2,2) circle (1pt);
      \draw[color=red] (-2.5,2) parabola bend (-0.25,0.8888) (2.5,2.5555) node[right,font=\tiny] {$P_2$};
  \end{tikzpicture}}
  \caption{Interpolating polynomials $P_0$ $P_1$ and $P_2$ based on the extremal eigenvalues only. They respect the stability condition (the stability area is coloured).}
  \label{fig:P012}
\end{figure}

In \cite{P2}, a method is presented to approximate $|\A|$ using interpolating polynomials $P_n$ of degree $n=0$, $1$ or $2$ (resp. denoted by $P_0$, $P_1$, $P_2$). The interpolation only focuses on the extremal eigenvalues $\lm$ and $\lM$, and adds a condition over one derivative in the $P_2$ case. Fig. \ref{fig:P012} depicts the graphs of the $P_0$, $P_1$ and $P_2$ polynomials. They all respect the stability condition (\ref{eq_stabi_poly}). For the targeted applications in which the orders of magnitude of the eigenvalues satisfy (\ref{eq:order_eigen}), it appears that the absolute values of the intermediate eigenvalues, which are close to zero, are not approximated accurately enough. This results in a quite poor approximation of $|\A|$ and gives rise to very diffusive schemes. These schemes are not accurate enough and will be discarded. In the following sections, we propose the construction of new polynomials which considerably improve the accuracy while maintaining the stability of the scheme.

\subsection[Approximation of |A| by high-degree polynomials]{Approximation of $|\A|$ by high-degree interpolating polynomials}
\label{sec:mypolyn}

We look for polynomials that provide accurate approximations of the absolute value function on the spectrum of the matrix, while maintaining the stability of the scheme when a phase appears or disappears. Such polynomials must satisfy the stability condition (\ref{eq_stabi_poly}) and avoid large oscillations near the eigenvalues. To meet the accuracy constraint, we need to consider high-degree polynomials. We consider polynomials interpolating the absolute value function in the interval $[-1,1]$. The general case can be deduced by applying the result to the matrix $\A / a_{\max}$, with $a_{\max}= \max_k |\lambda_k|$. Two approaches are considered: fixed interpolation and dynamic interpolation. Fixed interpolation means that the approximating polynomial does not depend on the eigenvalues and approximates the absolute value function in the whole range $[-1,1]$. Dynamic interpolation means that the approximating polynomial depends on the eigenvalues to be interpolated and focuses on the quality of the approximation near these eigenvalues. The second approach, although slightly more time-consuming, since it requires to re-construct the polynomial at each time-step and each cell-interface, will reveal to be more efficient.

\subsubsection{Fixed polynomial interpolation}

\begin{figure}
  \centering
  \subfigure[Even polynomials $P_{2p}$. \label{fig:even_polyn}] {\includegraphics[width=0.42\textwidth,trim=30 00 80 00,clip]{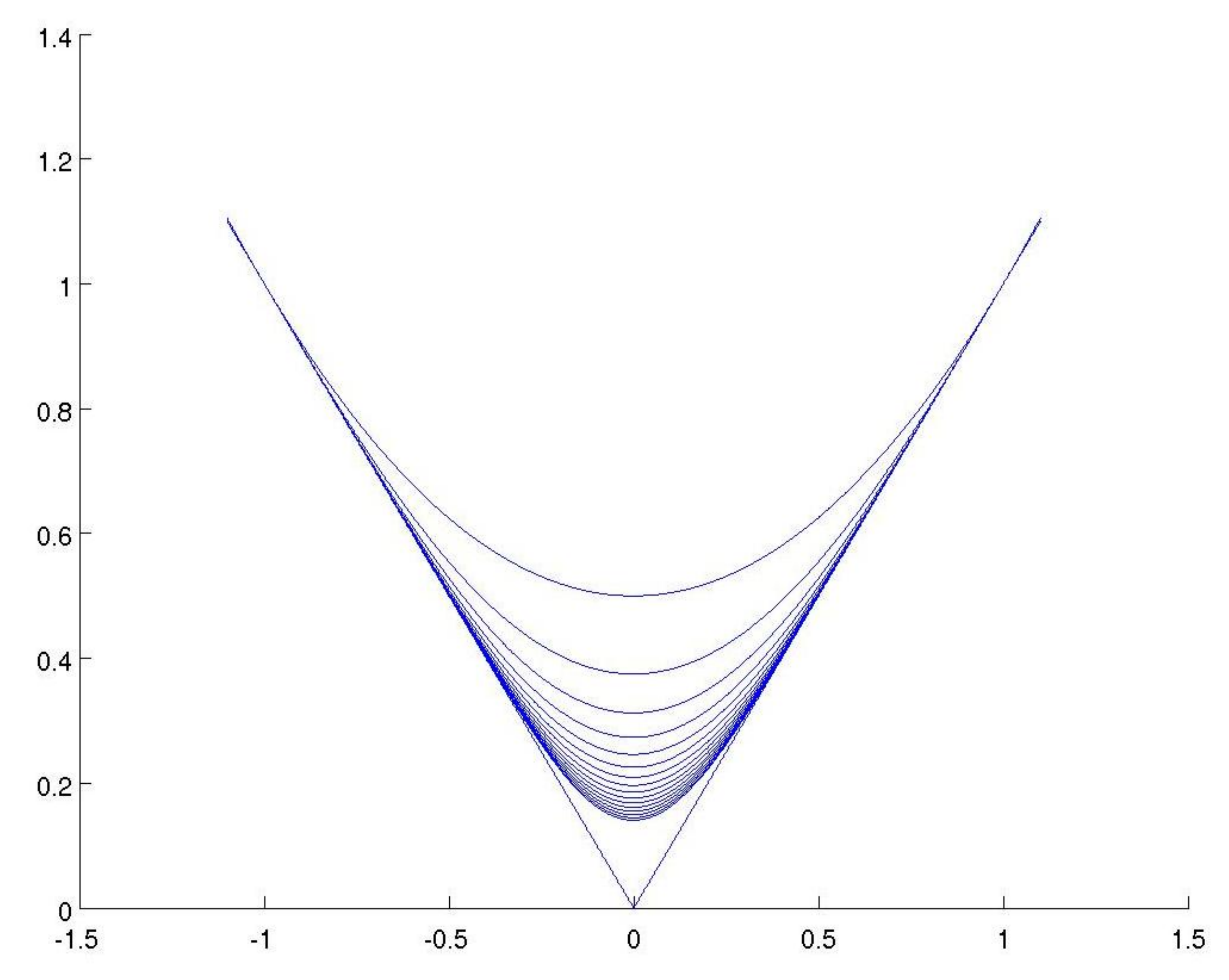}}
  \subfigure[$\Pds$ \label{fig:P17}]{\includegraphics[width=0.48\textwidth,trim=30 50 80 00,clip]{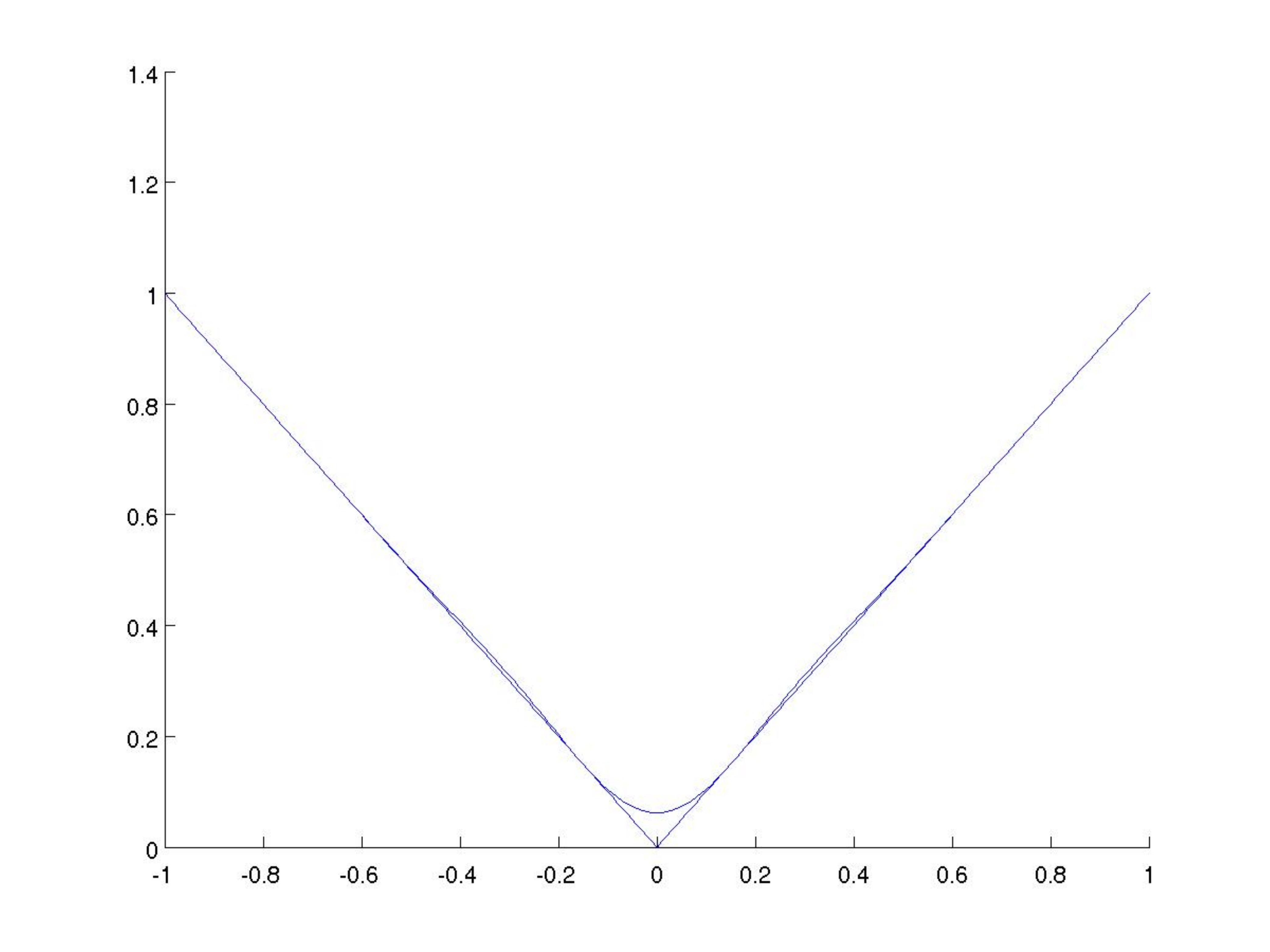}}
  \caption{(a) Even polynomials $P_{2p}$ interpolating the absolute value at the extremal points of the interval $[-1,1]$, for $p=1,\ldots,16$. (b) Even polynomials $\Pds$ interpolating intermediate points. }
  \label{fig:fixed_interp}
\end{figure}

\paragraph{Polynomials interpolating extremal points: the $P_{2p}$ polynomials.} A first idea is to construct even polynomials $P_{2p} = \sum_{k=0}^{p} a_k x^{2k}$ for $p\in \mathbb N$. $P_{2p}$ is constructed such that $P_{2p}(x) - |x|$ vanishes at $x=1$ as well as its derivatives up to the order $p$. The polynomial is then uniquely defined by: 
\begin{equation}
  \left\lbrace
  \begin{array}{l}
    P_{2p} \mbox{ is even, }\\
    P_{2p}(1) = 1, \\
    P'_{2p}(1) = 1, \\
    P_{2p}^{(j)}(1) = 0,\hspace{1em}j=2,...,p.
  \end{array}
  \right.
  \label{eq:even_polyn}  
\end{equation}
The coefficients $a_k$ are calculated once for all by solving a linear system. The larger the order of contact of $P_{2p}(x)-|x|$ with $0$, the better the approximation is. Fig. \ref{fig:even_polyn} displays $P_{2p}(x)$ for $p=1$ to $16$. The value $p=1$ corresponds to the interpolating polynomial $P_2$. Higher values of $p$ clearly provide better approximations of $|x|$. However, due to the inversion of the linear system, calculating the coefficients $a_k$ beyond $p=16$ presents numerical instabilities. The so-obtained accuracy is still not entirely satisfactory.

\paragraph{Even polynomials $\Pds$ interpolating intermediate points.} In this example, the even polynomial $P(x)= \sum_{k=0}^{p} a_k x^{2k}$ of degree $2p$ interpolates $|x|$ at a series of $m$ points $(x_i)_{1\leq i \leq m}$, $x_i \in [0,1]$, such that $P(x) - |x|$ has a contact of order $c_i$ with $0$ at $x_i$. The polynomial is determined by: 
\begin{equation}
  \left\lbrace
  \begin{array}{l}
    P_{2p} \mbox{ is even, }\\
    P(x_i) = |x_i| \\
    P'(x_i) = 1 \\
    P^{(j)}(x_i) = 0,\hspace{1em}j=2,...,c_i
  \end{array}
  \right.
  \label{eq:P17}  
\end{equation}
The degree of the polynomial is $2\sum_{i=1}^m (c_i+1)$. After several trials, it appeared that an almost optimal choice was obtained with the following parameters:
\begin{equation}
  \left\lbrace
  \begin{array}{l}
    p=17 \\
    x_1= 1   \hspace{1em} c_1= 7\\
    x_2=  t_1  \hspace{1em} c_2= 7\\
    x_3=  t_2  \hspace{1em} c_3= 1
  \end{array}
  \right.
  \label{eq:P17b}  
\end{equation}
where the $t_i$ are the two Tchebychev points on the interval $[0,1]$. The Tchebychev points give minimal oscillations for an interpolating polynomial, and have the following expression for an interval $[a,b]$ divided in $n$ points:
\begin{equation}
  t_k = -\frac{b-a}{2} cos\frac{(2k+1)\pi}{2(n+1)} + \frac{a+b}{2}
  \label{eq:tchebychev}
\end{equation}
We will refer to this polynomial as $\Pds$, for "High-Degree Fixed" polynomial. Its coefficients are calculated once for all by solving the linear system (\ref{eq:P17}). The numerical values of the coefficients are given in appendix \ref{sec:coeffp17}. As we have 
$$ \min \limits_{x \in [0,1]}( \Pds(x) - |x| ) \sim -10^{-13}, $$ 
we add a constant equal to $10^{-10}$ so that the polynomial remains greater than the absolute value function. The so-obtained polynomial respects the stability condition. The graph of $\Pds$ is given in Fig. \ref{fig:even_polyn} (b). The approximation of the absolute value function by $\Pds$ is improved. However, the absolute value of the intermediate eigenvalues, those which have a magnitude close to $0$, remain inaccurately approximated and the scheme remains too diffusive.

\subsubsection{Dynamic polynomial interpolation}
\label{subsec:pfp}

As seen is the previous section, to improve accuracy, it is necessary to take into account the intermediate eigenvalues. The resulting polynomial depends on the eigenvalues, and motivates the terminology 'dynamic interpolation'.

One of the reasons for the large oscillations of the interpolation polynomial $\Pexact$ is the presence of a cluster of intermediate eigenvalues near $0$, which are very close to each other and which are $4$ orders of magnitude smaller than the extremal eigenvalues (see (\ref{eq:order_eigen})). In \cite{sedes}, an approximate Roe matrix is generated by treating this cluster of eigenvalues like a single eigenvalue (the largest of them). The method of \cite{sedes} provides comparable results to the usual Roe method in standard situations. Inspired by this work, we generate an approximate polynomial which interpolates the extremal eigenvalues $\lm$, $\lM$ and only one of the intermediate eigenvalues, the largest one, denoted by $\lintM$ (as well as its opposite $-\lintM$ for symmetry reasons). It thus avoids the interpolation of the cluster of very close intermediate eigenvalues. Conditions on derivatives are added so that the stability condition is respected locally around the eigenvalues. The polynomial is a Hermite interpolation polynomial constructed in Newton's basis, and is calculated at each time-step and for each cell-interface. The computation does not break down at phase appearance or disappearance. Indeed, the collapse of eigenvalues only concerns intermediate ones. The design of the polynomial considers already only one of them and it does not matter how many of them are distinct. 

\begin{figure}
  \centering
  \subfigure[ $\lintM= 10^{-3}$ \label{fig:4pts1}]{\includegraphics[width=0.43\textwidth,trim=30 0 80 00,clip]{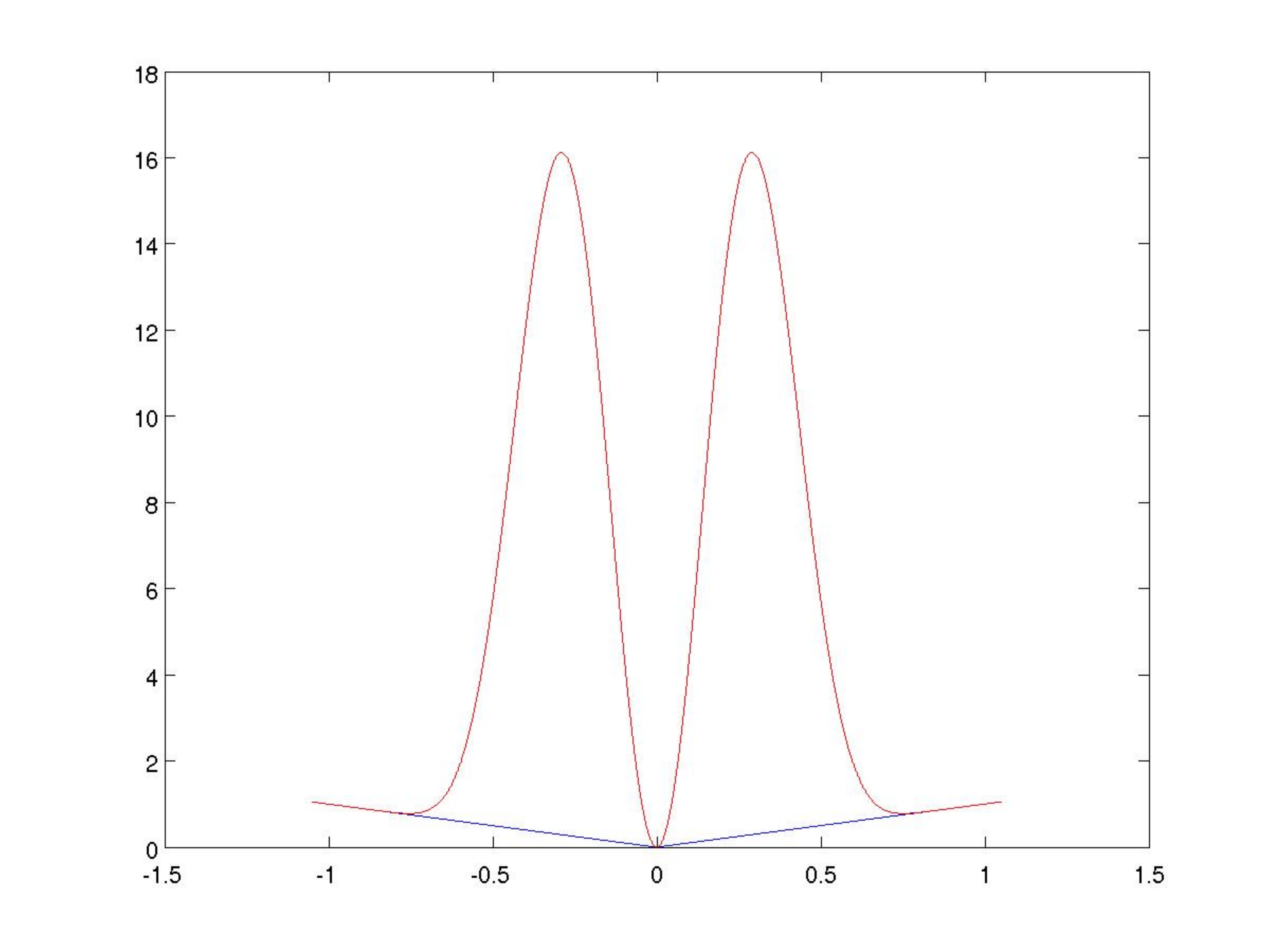}}
  \subfigure[Zoom on $x=0$\label{fig:4pts2}]{\includegraphics[width=0.43\textwidth,trim=30 0 80 00,clip]{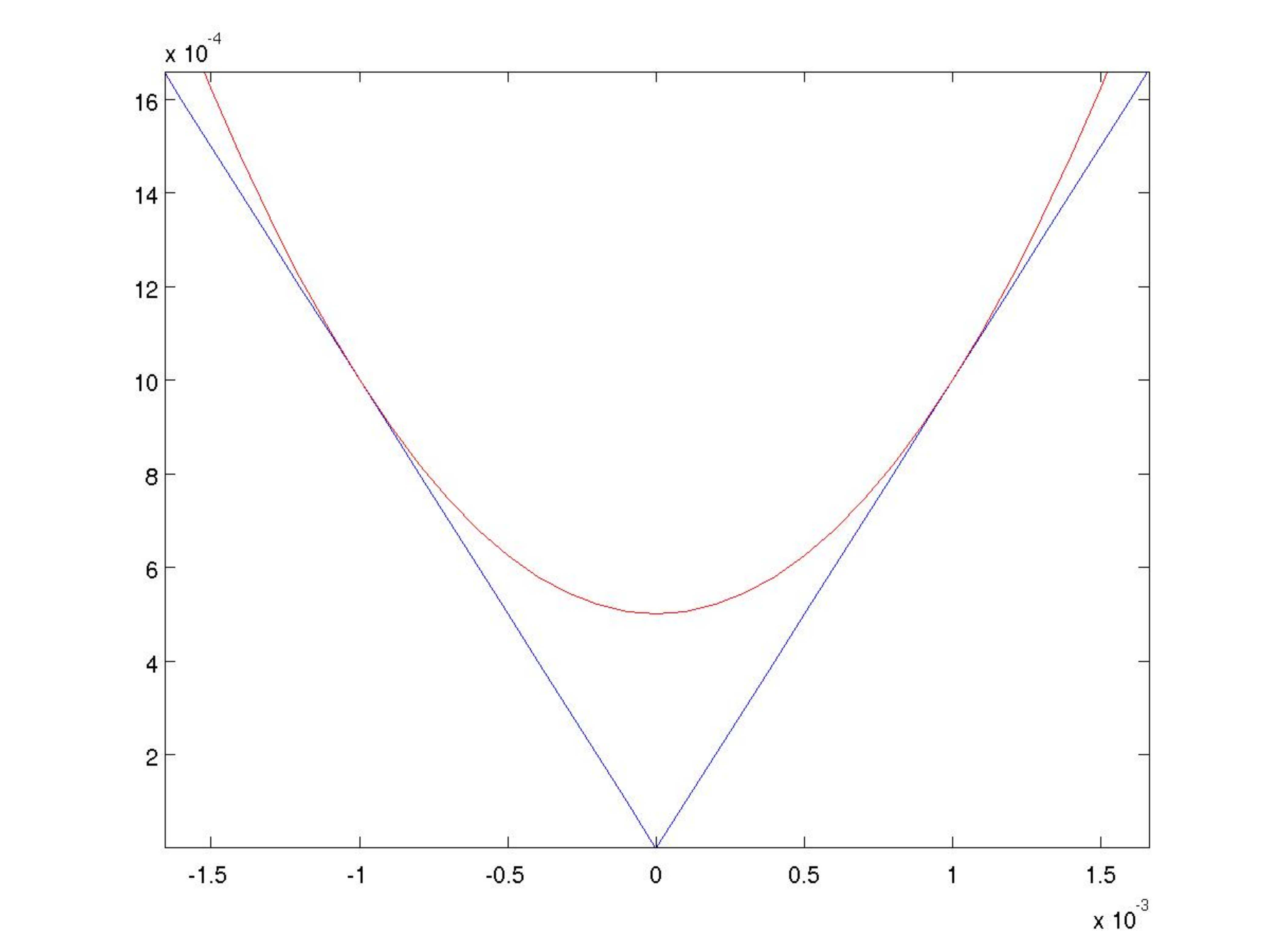}}
  \caption{Dynamic interpolating polynomial $\pfp$ based on the interpolation of the extremal eigenvalues and the largest intermediate eigenvalue (left). There are no eigenvalues in the region of the large oscillations in the median regions and the stability condition (\ref{eq_stabi_poly}) is respected locally about the eigenvalues. The right figure shows a blow-up of the left picture near $0$}
  \label{fig:p4points}
\end{figure}

We will call this polynomial $\pfp$, for "High-Degree Dynamic" polynomial. It verifies the following conditions : 
\begin{eqnarray*}
& & \pfp(\lm)=|\lm| , \qquad \pfp(\lM)=|\lM|, \\
& & \pfp(\pm\lintM)=|\lintM|, \qquad  \pfp'(\lm)= -1, \\
& &\pfp'(\lM)= 1, \qquad \pfp'(-\lintM)=- 1, \\
& & \pfp'(\lintM)= 1, \qquad  \pfp^{(j)}(\lm)=\pfp^{(j)}(\lM)= 0,\hspace{1em}j=2,...,10. 
\end{eqnarray*}
As fig. \ref{fig:4pts1} (left) shows, the contact between the polynomial and the absolute value is very good in the neighborhood of the extremal eigenvalues, as the first derivative is set to $\pm 1$ (the derivative of the absolute value) and the other derivatives are set to zero. The large oscillation between the extremal eigenvalues and the intermediate eigenvalues is not a problem because there exist no eigenvalues in this region. This allows us to approximate $|x|$ near $x=0$ very accurately. Fig. \ref{fig:4pts2} (right) shows a blow-up close to zero. We notice that the stability condition (\ref{eq_stabi_poly}) is indeed verified for all intermediate eigenvalues, as we do have $\pfp(\la) \geq |\la|$ for all intermediate eigenvalues.

\bigskip

We will test the behavior of the $\Pds$ and $\pfp$ polynomial schemes in the numerical section \ref{sec:numres}. We will see that numerical difficulties are reduced but some positivity problems remain. In the following part, we develop a method to specifically treat the positivity problems. It will rely, among others, on the possibility of tuning the amount of diffusion in the $\pfp$ polynomial. But first, let us detail an other way to compute $|\A|$ without using the eigenstructure of the matrix. Based on the same principle as the polynomial solvers, the method uses the hyperbolic tangent function.

\subsection{Approximation of $|\A|$ by means of the hyperbolic tangent}
\label{sec:tanh}

In this section, we present an alternative to the use of the polynomial schemes. We recall that the goal is to compute the absolute value matrix $|\A|$ without using the eigenvector decomposition of $\A$. We introduce the following approximation $\Phi(x)$ of the absolute value function $|x|$: 
\begin{equation}
  \Phi(x) = \tau + (1-\tau) \, x \,  \tanh(\frac{x}{\tau}) \,  \text{cotanh}(\frac{1}{\tau}).
  \label{eq:phix}
\end{equation}
with 
$$ \tanh(x)=\frac {e^x-e^{-x}} {e^x+e^{-x}} \, , \quad \text{cotanh}(x)=\frac{1}{\tanh(x)},$$ 
and $\tau>0$ is a parameter.

\begin{figure}[hb]
  \centering
  \subfigure[$\tau=0.1$]{\includegraphics[width=0.38\textwidth,trim=30 50 80 60,clip]{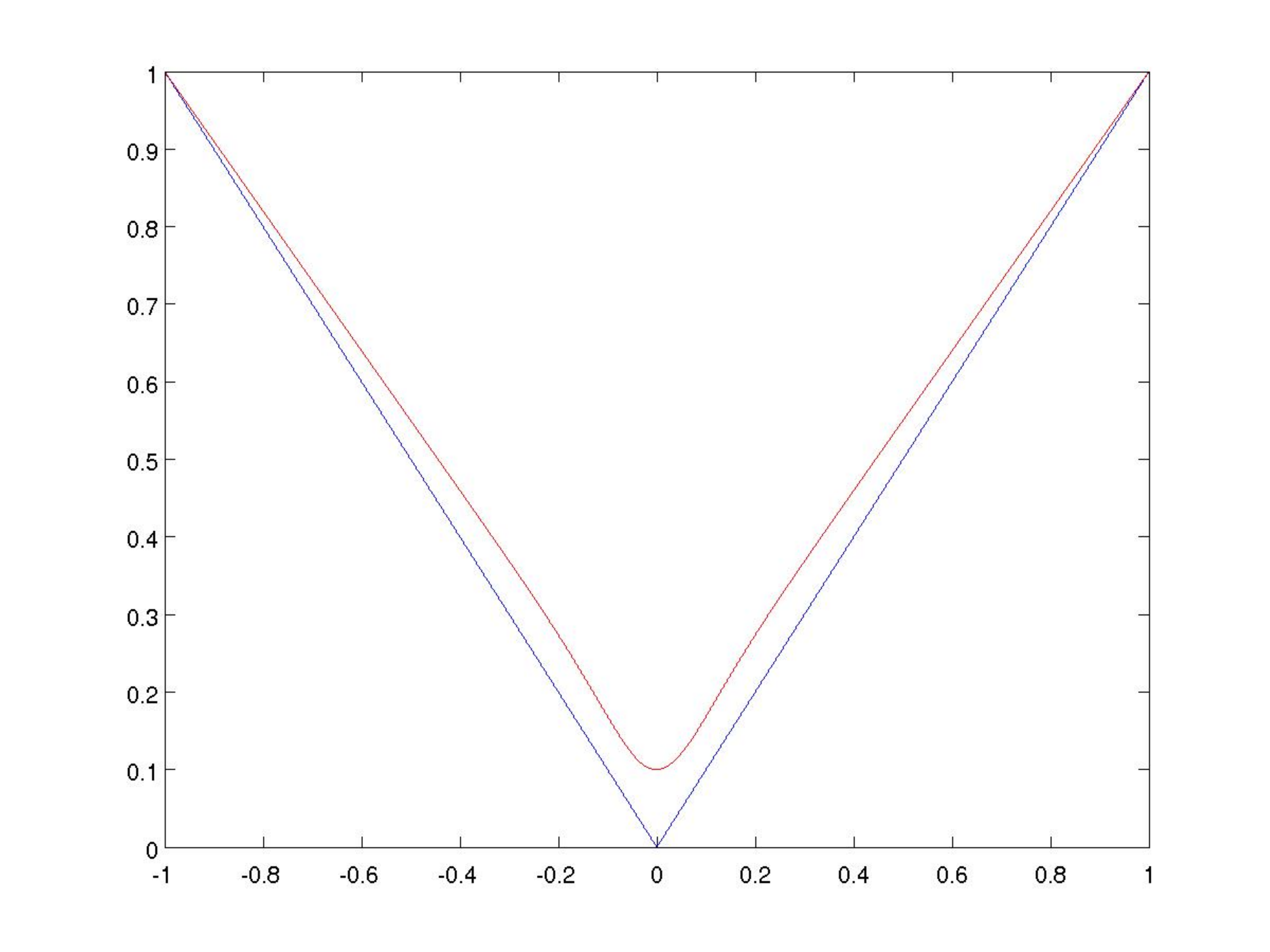}} 	
  \subfigure[$\tau=0.001$]{\includegraphics[width=0.38\textwidth,trim=30 50 80 60,clip]{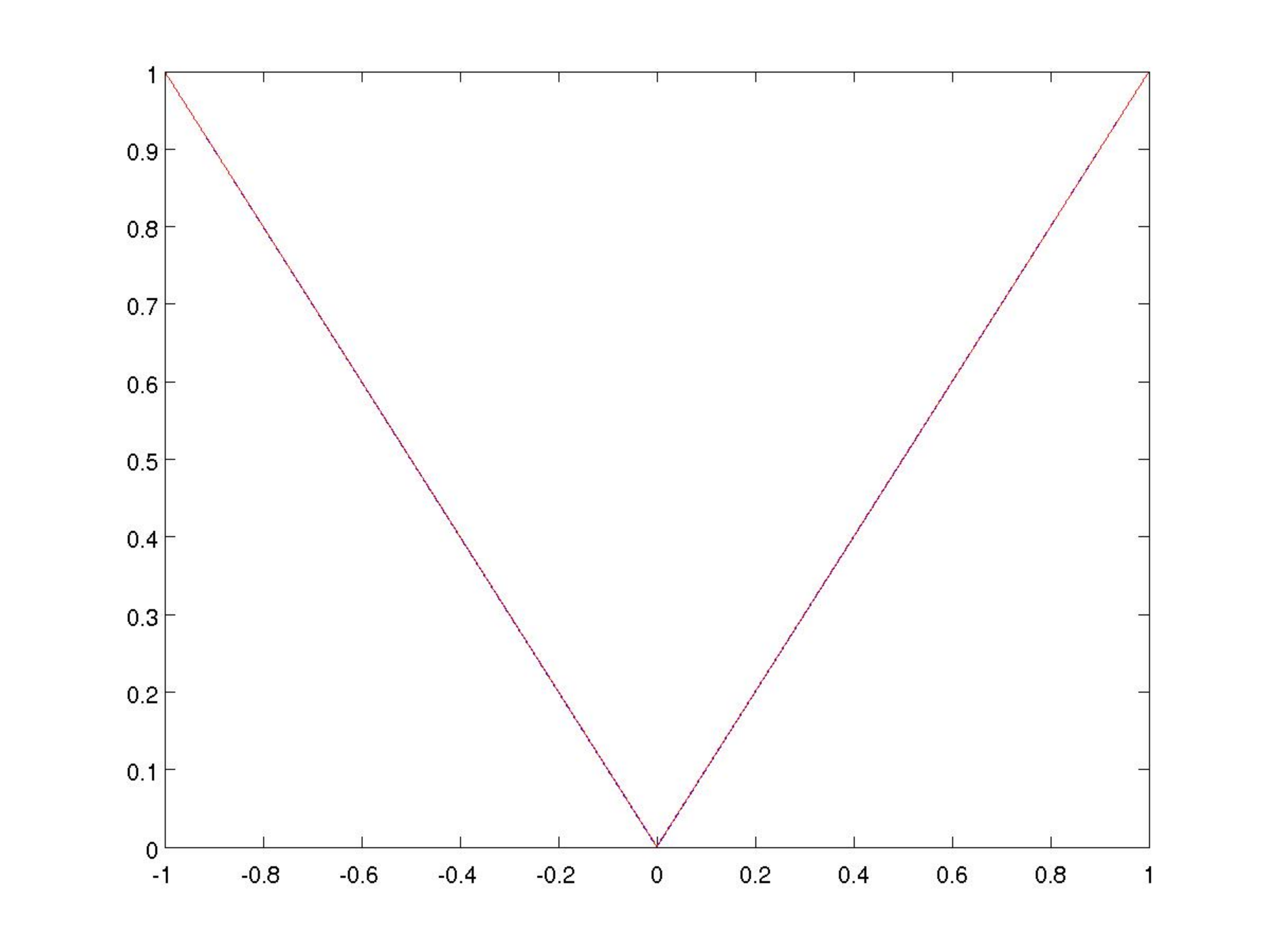}}
  \caption{The function $\Phi(x)$ (in red) as a function of $x \in [-1,1]$: the parameter $\tau$ controls the accuracy of the approximation of $|x|$ (in blue) by $\Phi(x)$.}
  \label{fig:tanhphi}
\end{figure}

As in section \ref{sec:mypolyn}, we will normalize the matrix $\A$ by the largest absolute value of the eigenvalues and study the function $\Phi$ only in the interval $[-1,1]$. We have $\Phi(1) = 1$ and $\Phi(x) \geq |x|$, for all $x \in [-1,1]$. Furthermore, it is easy to realize that $|\tanh(\frac{x}{\tau})- S\text{ign}(x)| \to 0$, and consequently, that $\big|\Phi(x) - |x|\, \big| \to 0$  when $\tau \to 0$, uniformly for $x \in [-1,1]$, where $S\text{ign}(x)$ denotes the sign function. Therefore, $\Phi$ obeys the stability condition (\ref{eq_stabi_poly}) (see also fig. \ref{graph:stability}) and is an approximation of $|x|$ with a controlled accuracy. The graph of $\Phi$ is represented on fig. \ref{fig:tanhphi} for different values of the parameter $\tau$. 

The practical choice of $\tau$ is performed as follows. Our goal is to approximate closely all the non-zero eigenvalues of $\A$ including the smallest. Consequently, we must have $\tau < \min\limits_{\li\neq0} |\li|= \lambda_{s}$.  Tab. \ref{tab:tanherror} presents the maximum error on the interval $[\lambda_{s},1]$ between $|x|$ and $\Phi(x)$ for different values of $\tau$ and for $\lambda_s = 10^{-4}$. The choice of $\tau = \lambda_s/10$ gives already quite good accuracy. Nevertheless, if we want to add some diffusion, it is possible to reduce the accuracy by increasing the parameter $\tau$.

\begin{table}[h!]
  \centering
  \begin{tabular}{|c|c|}
    \hline 
    $\tau$ & $ \max\limits_{x\in [\lambda_{s},1]} |\Phi(x) - |x| |$\\ \hline 
    $\frac{\lambda_s}{10}=10^{-5}$ & $9.998 \times 10^{-6}$ \\
    $\frac{\lambda_s}{100}=10^{-6}$ & $9.998 \times 10^{-7}$ \\
    $\frac{\lambda_s}{1000}=10^{-7}$ & $9.999 \times 10^{-8}$ \\
    \hline
  \end{tabular}
\caption{Accuracy of the approximation of the absolute value function depending on $\tau$}
  \label{tab:tanherror}
\end{table}

We now present how we compute the hyperbolic tangent of the matrix $\A$ without using the eigenvectors of the matrix. We note that the scalar hyperbolic tangent function $x \to \tanh(\alpha x)$ (where $\alpha \in \mathbb R$ is a constant) satisfies the following differential equation:
\begin{equation}
  \frac{d} {d x} \tanh(\alpha x) = \alpha ( 1-\tanh(\alpha x)^2).
\end{equation}
Therefore, we solve the matrix differential equation: :
\begin{equation}
  \left\{
  \begin{aligned}
    \frac{d \X(\zeta)}{d\zeta} &= \A (\mathbb{I}-\X(\zeta)^2) \\
    \X(0) &=0. 
    \label{eq:ED}
  \end{aligned} 
  \right.
\end{equation}
which yields $\X(\zeta) = \tanh(\zeta\A)$ with $\zeta\in \mathbb R$. 
We solve this differential equation by means of an iterative implicit method. An explicit method based on a fourth order Runge-Kutta method can also be used but the computational cost is higher. The scheme is written:
\begin{equation}
  \left\{
  \begin{aligned}
 &   \X^k \approx \X(\zeta_k)\\
 &   \X(0) = 0 \\
  &  \X^{k+1}= \X^k + h \A ( \Id - (\X^{k+1})^2 )
  \end{aligned}
  \right.
\end{equation}
\noindent where $h$ is the iteration step. We use the Newton method to find $\X^{k+1}$ at each iteration.  The number of steps needed to prevent the algorithm from diverging can be set to a constant value $N_1$. Each iteration contains another loop of maximum $N_2$ iterations to find $\X^{k+1}$ by the Newton method. We have chosen $N_1=100$ and $N_2=40$.

\section{Numerical treatment of positivity losses}
\label{sec:pos}

Positivity problems tend to appear during the simulation of phase transitions. We will first review previously developed positive schemes. We will then propose an other method to solve the positivity problems, method complying with the constraints mentioned in the introduction : no analytical expression of the eigen-elements are available, the computation of the eigenvectors should not be used during phase transitions, and the method has to be compatible with an implicit scheme or large time-steps. We will give the general principle of the method and then the features of its implementation.

\subsection{Previous works on positive numerical schemes}

A positive scheme for the two-fluid two-phase flow model has been proposed in \cite{coquel1997methode}. The explicit scheme introduces a splitting in the resolution of the bifluid model. The first step (hydrodynamic step) solves separately two uncoupled full Euler systems, for each phase, by means of a kinetic solver, for stability reasons. The non conservative terms in $\partial_x \alpha$ are reformulated and included in the source terms. A second step enforces the equality of the pressures and allows to compute directly the void fraction and the pressure. This scheme preserves the positivity of all thermodynamics variables under a CFL-like condition.

As few works concern the resolution of the two-fluid two-phase flow model specifically, let us also mention the positive numerical schemes that have been designed for Euler equations, or gas dynamics equations.

Einfeldt and al. \cite{einfeldt1991godunov} consider the HLLE solver for the Euler equations. HLLE is positively conservative, but less accurate than the Roe scheme. Anti-diffusion parameters are introduced in the HLLEM scheme to take out excessive dissipation.
In \cite{perthame1996positivity}, Perthame and Shu show that the Lax-Friedrichs scheme is positivity preserving, which echoes the general observation that the more diffusive a scheme is, the more robust it is, robustness including here positivity preservation. 

For the sake of accuracy, other works introduce second order schemes. The positivity constraint is ensured by limiting techniques. In \cite{jameson1995positive}, the symmetric limited positive scheme (SLIP) conserves the positivity thanks to the use of a limited diffusive flux which makes the scheme local extremum diminishing (LED). This property is stronger than the total variation diminishing (TVD) property proposed by Harten \cite{harten1983high} (if the scheme is LED, then it is TVD, LED and TVD being equivalent in one dimension), and ensure that a local maximum cannot increase, and a local minimum cannot decrease. Thus, if the solution is positive at one moment, than the global minimum is positive and cannot decrease and become negative. This SLIP scheme can be applied to the Roe scheme for a system of conservation laws. The construction of the scheme requires the computation of the eigenvectors.

The Maximum Limited Gradient reconstruction technique \cite{batten1996positively}, also gives a second order positivity preserving method. In \cite{liu2005positive}, Liu and Lax propose a family of second order positive schemes for multi-dimensional hyperbolic systems of conservation laws, using a limiter in the numerical flux. The eigenvectors and eigenvalues of the Roe matrix have to be evaluated explicitly to construct the scheme. The second order central scheme described in \cite{jiang1996nonoscillatory} is based on a predictor corrector method and is positive due to the scalar maximum principle. An other positive scheme based on flux limiters can be found in \cite{berzins1995positive}.

Other methods designed to address the lack of positivity have also been based on a modification of the Roe scheme. Dubroca in \cite{dubroca1999solveur} and Gallice in \cite{gallice} propose extensions of the Roe's solver for the Euler equations on the one hand and for systems of magnetohydrodynamics on the other hand. In \cite{einfeldt1991godunov}, Einfeldt and al. concluded that there is no positively conservative Roe matrix, but also specified that this statement only applies to Roe's matrices based on Jacobian matrices. These works introduce the derivative of the pressure in the direction of the fluid velocity in the Roe matrix. This decomposition allows to introduce parameters that can be chosen so that the solver becomes positively conservative. The demonstrations of the positivity of the two methods are based on the eigenvalues and eigenvectors of the Roe matrix, that can be calculated analytically for the considered systems.

An noticeable point in \cite{gallice} is the establishment of a link between diffusion and positivity. Gallice introduces parameters in the Roe matrix as a way to exactly tune the dissipation so that the scheme becomes positive. Our work is based on the same idea:  finding the right amount of diffusion so that the positivity problem can be solved. 

Up to now, most of the methods available in the literature are based on the construction of specific schemes which are designed to prevent the positivity problems. A different, less developed strategy is to treat the positivity problem when it appears rather than trying to prevent it. Such a strategy is proposed by Romate in \cite{romate1998approximate}. Romate remarks that the HLL scheme \cite{einfeldt1988godunov} is positive under a CFL condition but too diffusive for practical use as such, while the Roe scheme is accurate but is not positive. He presents a combination of the two schemes. The Roe flux is used until the appearance of a positivity loss in an adjacent cell. If such a problem occurs, the time-step computation is restarted using the the HLL flux. The newly computed cell value will therefore be positive.

The method presented below is inspired from Romate's strategy \cite{romate1998approximate}.  Rather than trying to prevent positivity problems from appearing, it consists in applying a special treatment when such problems appear. It is also inspired from Gallice \cite{gallice} in that the treatment consists in increasing the numerical diffusion in some way.

\subsection{Description of the method}

The method is inspired from \cite{gallice} where positive Roe schemes for the gas dynamics and MHD equations are proposed. This work is based on the fine tuning of the numerical diffusion in order to make the scheme positive. The availability of analytic expressions for the eigenvalues of the Roe matrix is a key ingredient in the demonstration of the positivity preserving property. In our case, the two-phase model is too complex to allow for analytic expressions of the eigenvalues. Thus, we will develop the idea in a different way. We increase the numerical diffusion where positivity losses are detected. To this purpose, we use the polynomial scheme based on the $\pfp$ polynomial just described. Indeed, it provides an easy way to adjust the numerical diffusion. We stress that, by contrast to \cite{gallice}, we have no proof that positivity is preserved, but only numerical evidence that robustness against loss of positivity is enhanced. 

To introduce numerical diffusion within the $\pfp$ polynomial, we just reduce the accuracy of the interpolation of the intermediate eigenvalues, thus making the scheme more diffusive. Instead of the points $(\pm\lintM,|\lintM|)$, we interpolate the points $(\pm\lintM,|D \,\lintM|)$, where  
$D\geq 1$ is a diffusion coefficient which is chosen to maintain the condition  $P(\lintM)>|\lintM|$, in agreement with the stability condition (\ref{eq_stabi_poly}).

When no positivity problem appears, the code is normally run with the $\pfp$ polynomial associated to a diffusion coefficient $D=1$. If, after a time-step, a positivity problem is detected in some cell, the computation of the time-step is restarted using a new $\pfp$ polynomial using $D=10$ in the adjacent cell interfaces. If the positivity problem remains, we further increase $D$ (the precise algorithm is given below). If finally, the maximal value of $D$ is reached (beyond which $D|\lintM| > \max(|\lm|,\lM)$, which is forbidden by the stability condition (\ref{eq_stabi_poly})), then, we stay with this value of $D$ and reduce the time step. 
The details of the algorithm are given below. 

In the numerical section \ref{sec:numres}, we will see that the combination of the $\pfp$ polynomial scheme and the present treatment of positivity problems leads to significant improvements. The robustness and reliability of two-phase flow codes in situations of phase appearance and disappearance is greatly enhanced.

\subsection{Implementation}
\label{sec:algopos}

We use the $\pfp$ polynomial scheme to adapt the diffusion by the means of the coefficient $D$ so that we interpolate $D\,|\lintM|$ instead of $|\lintM|$, the largest intermediate eigenvalue. If $D>1$, the diffusion is increased. A coefficient $c_i$ is attributed to each cell to take inventory of the occurrence of positivity problems within a given time-step. The treatment proceeds according to the following algorithm, which describes a time-step advance. We will call this method $\pfppos$.

\begin{enumerate}[a -]
\item At the beginning of the time-step, the counter $c_i=0$ on all cells: no positivity problems has occurred yet.
\item On all interfaces, compute $|A|$ with $\pfp$ and $D=1$.
\item Solve the time-step.
\item Loop around all the cells to check for problems. For all cells $i$ where a positivity problem or a convergence problem during the computation of the variables of states (pressure, enthalpy) has occurred, increase the counter $c_i$ by $1$.
\item Compute $|A|$ again with an increased diffusion on the interfaces where at least one of the neighboring cell is such that $c_i \neq 0$. In order to increase progressively the diffusion, we use $D=10 c_i^3$. If $D\,|\lintM| > \max(|\lm|,\lM)$, we set $D=\max(|\lm|,\lM)/|\lintM|$ in order to remain in the domain of validity of the stability condition (\ref{eq_stabi_poly}).
\item Solve the time-step again and iterate if necessary until all problems are solved. We set up initially a maximal number of iterations. If this number is reached, we reduce the time-step $\Delta t$ and restart the computation of the time-step. 
\end{enumerate}

This method allows to overcome most positivity issues, but also problems which may occur in the computation of the variables of states (pressure, enthalpy). It is very important to note that, as diffusivity is added very locally, this method has no negative repercussion in terms of global accuracy, which is preserved. Let us also note that, in \cite{gallice}, there is no analytic expression of the parameters introduced to correct the system matrix so that the scheme is positive. These parameters have to be "large enough" to ensure the positivity, and their value is also obtained by an iterative procedure, as in our method.

\section{Numerical results}
\label{sec:numres}

We present several test-cases in one and two dimensions. The Ransom faucet test-case will first allow us to compare the accuracy of the different polynomial schemes. The boiling channel in the saturated case will then highlight the improvement brought by polynomial schemes in a situation of phase appearance or disappearance. We will then show more difficult test-cases where the positivity treatment is required : the boiling channel in the subcooled case, and the two-dimensional tee-junction test-case. In all test-cases, the water-and-steam equation of state will be used. The International Association for the Properties of Water and Steam (IAPWS) provides internationally accepted formulations for the properties of steam and water. In all test cases, the used model is that of section \ref{sec:setfm}, and the right-hand sides will be specified precisely.

\subsection{Ransom faucet}

This non-stationary one-dimensional test-case was proposed by Ransom in \cite{ransom1987nbt}. It considers the flow of a water column at the outlet of a faucet opening out into a vertical enclosure containing air. The considered tube is $12$ $m$ high, and the inlet velocity is $10$ $m/s$ whereas the air is at rest. The ratio of the sections of the nozzle and of the enclosure is such that the integrated void fraction over the section is equal to 0.2. In this configuration, a striction phenomenon of the jet is observed due to the effect of gravity. Indeed, if we make the assumption that the jet remains coherent (no tear-off of liquid in the form of drops, no penetration of air into jet), the acceleration of the liquid due to gravity necessarily results in a narrowing of the cross section of passage of the liquid, by conservation of the flowrate. Moreover, as the initial conditions correspond to the solution which would be obtained in the absence of gravity (therefore $\av=0.2$ everywhere), a void fraction discontinuity is propagated from the inlet section to the outlet section as from the initial time.

This test-case allows to evaluate the accuracy of a scheme, the amount of numerical diffusion being visible on the void fraction front. The velocity of the void fraction wave has to be correctly captured. The model is the \setfm presented in section \ref{sec:setfm}. The only source terms are the interfacial pressure term (\ref{term:bestion}) with $\delta_0 = 1.1$, and the gravity with $g=9.81 m/s^2$. At the inlet, the following values are fixed:  $u_v =  0.0$ m/s, $u_\ell = 10.0$ m/s, $h_v = 324.594$ kJ/kg, $h_\ell = 209.283$ kJ/kg, and $\alpha_v=0.2$. At the outlet, the pressure is fixed: $p_{outlet} = 10^5 \text{ Pa}$. The computational method is explicit. We used a one-dimensional mesh with 100 cells. The maximum time is $t = 0.6$ $s$.

\begin{figure}[t]
  \centering
  \subfigure[Void fraction]{\includegraphics[width=0.48\textwidth,trim=30 50 0 80,clip]{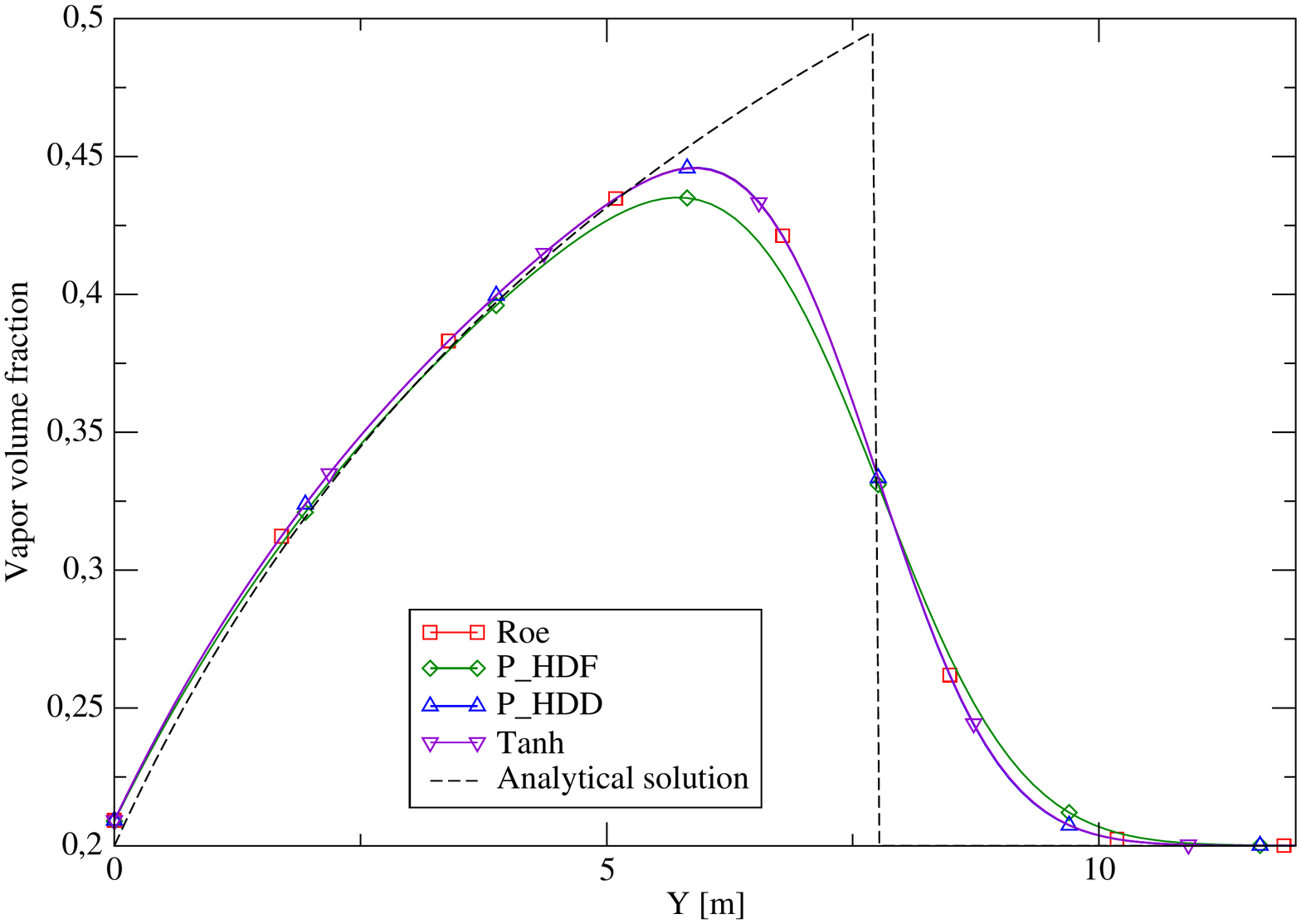}}
  \subfigure[Pressure]{\includegraphics[width=0.48\textwidth,trim=30 50 0 80,clip]{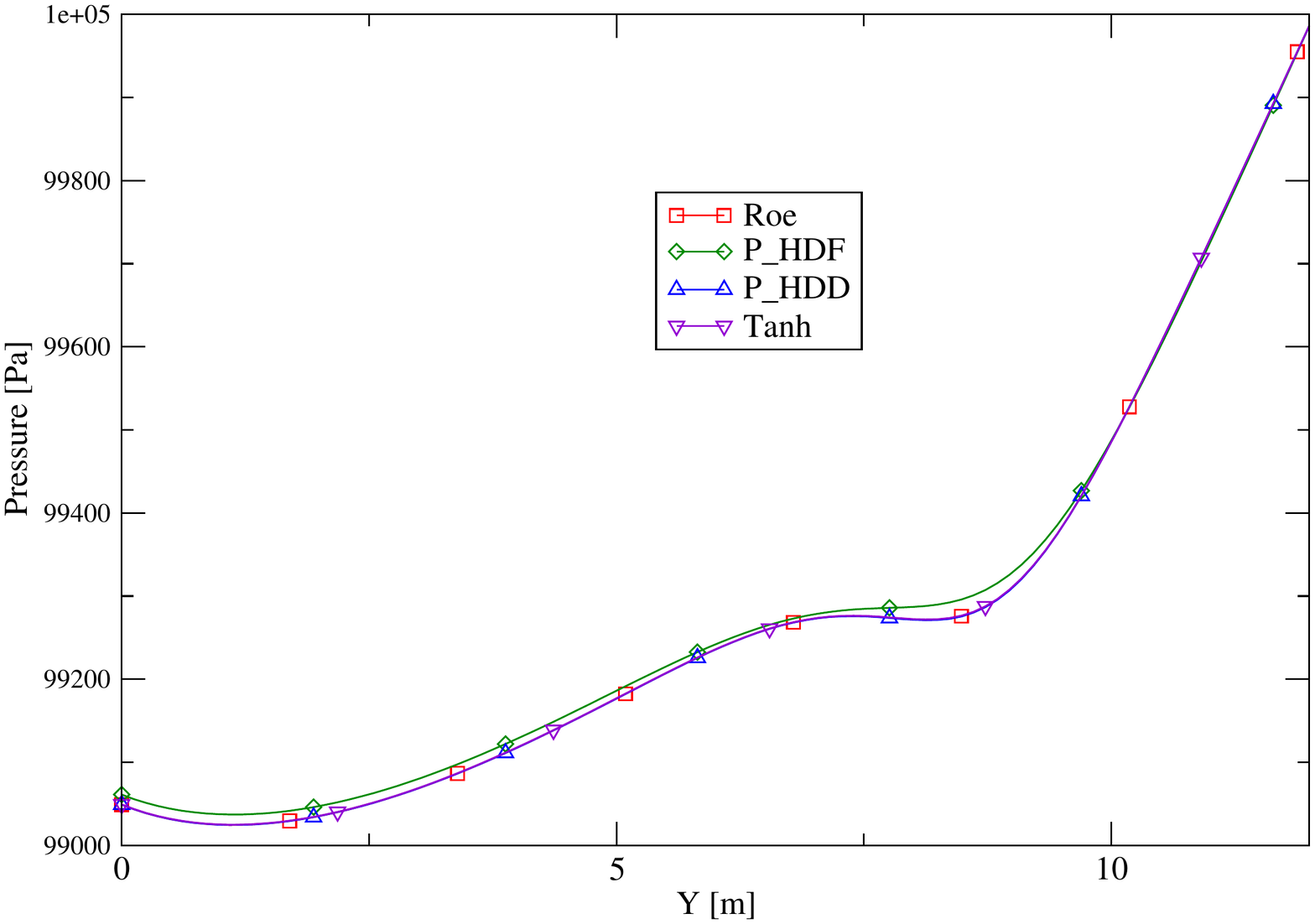}}

  \subfigure[Liquid velocity]{\includegraphics[width=0.48\textwidth,trim=30 50 0 80,clip]{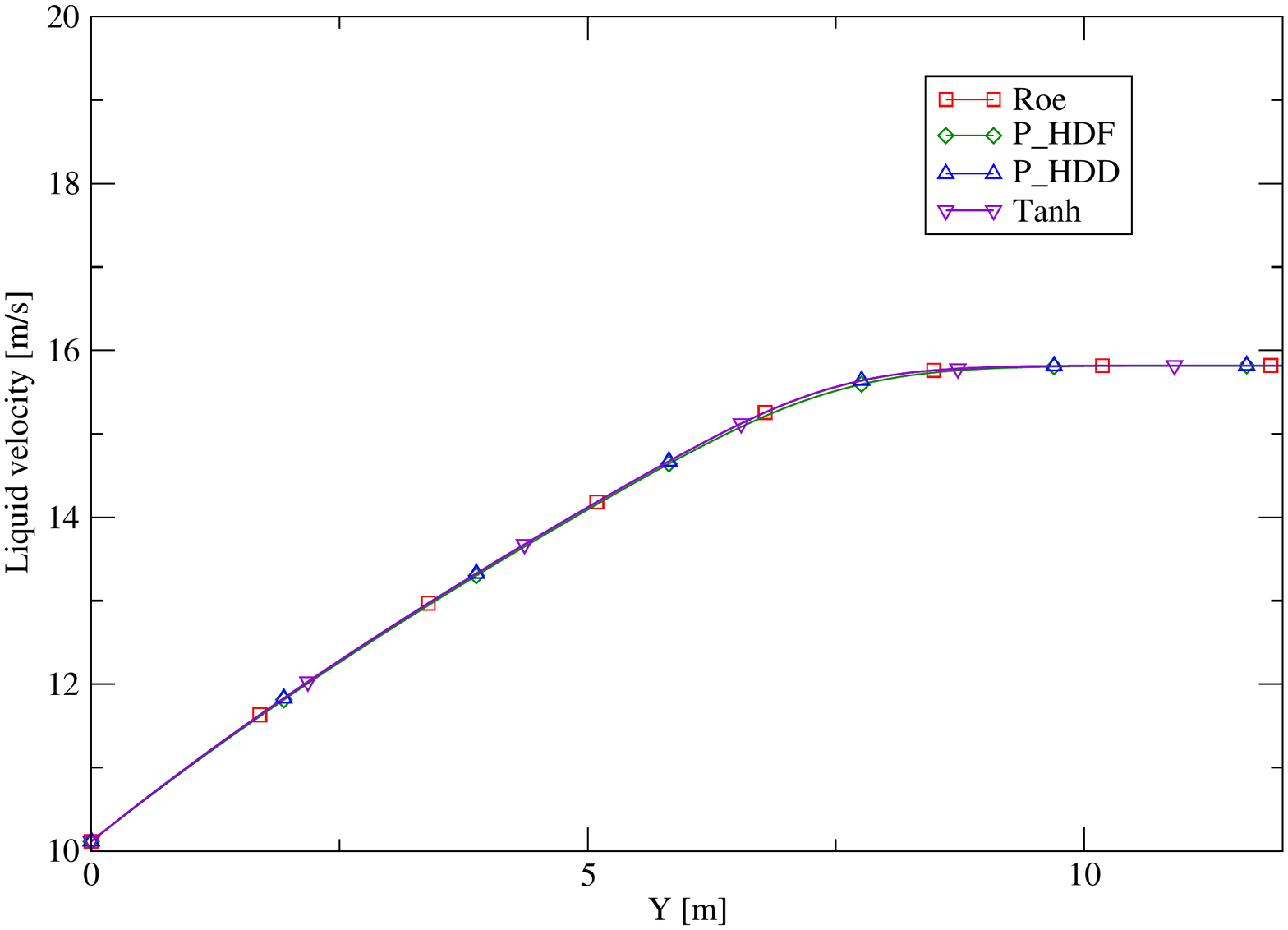}}
  \subfigure[Vapor velocity]{\includegraphics[width=0.48\textwidth,trim=30 50 0 80,clip]{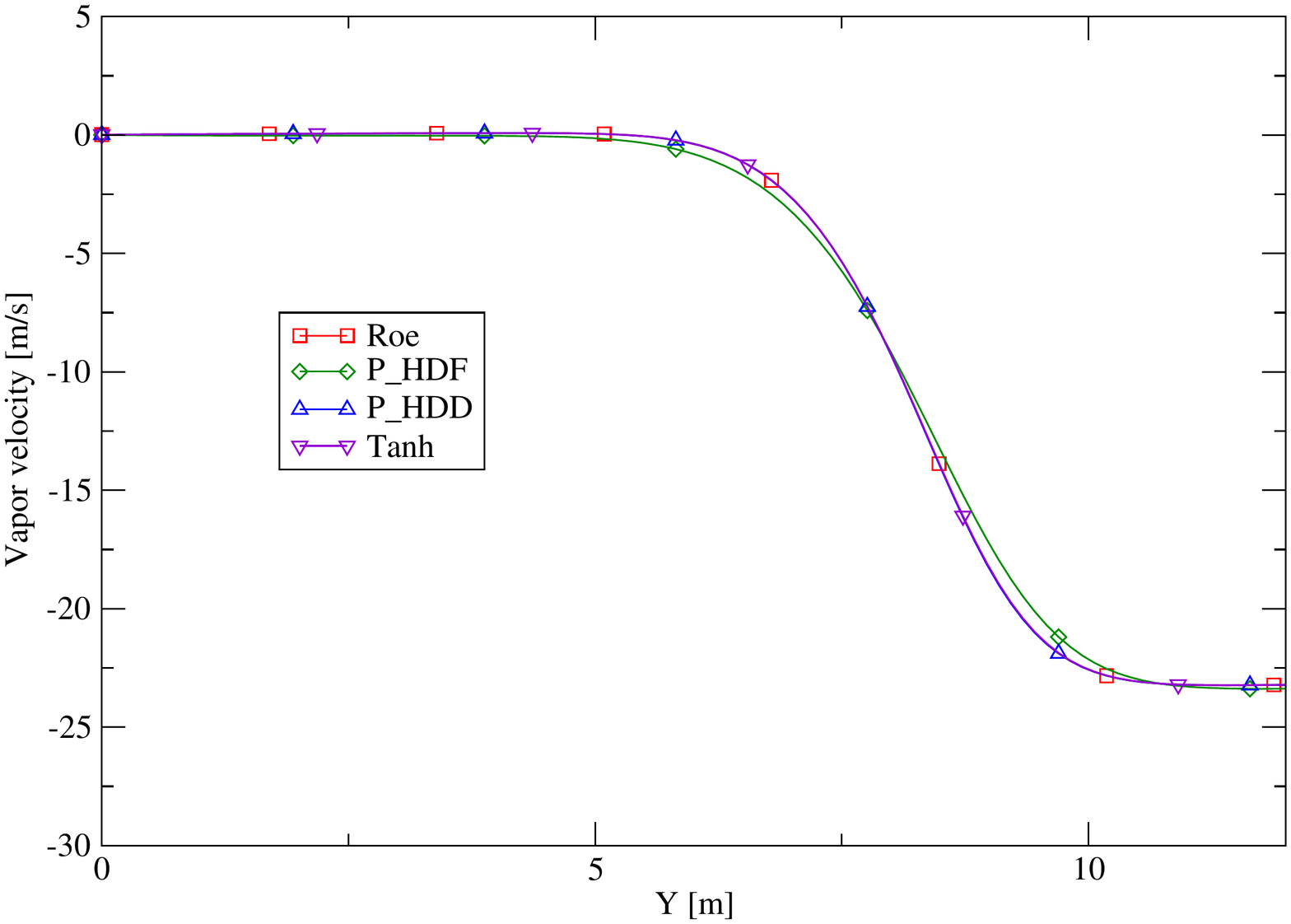}}
  \caption{Ransom faucet with 100 cells. Void fraction (a), pressure (b), liquid velocity (c) and vapor enthalpy (d) at time $t=0.6$ s, as functions of the height in the column. The Roe scheme (red squares), $\Pds$ (green diamonds) and $\pfp$ (blue triangles) polynomial solvers, and the $Tanh$ (violet triangles) method are compared. The analytical solution for the void fraction is shown in black dashed line on figure (a). }
  \label{fig:rf1_polyn}
\end{figure}

Fig. \ref{fig:rf1_polyn} represents the profiles of the volume fraction, pressure and velocities, for the $\Pds$ and $\pfp$ polynomial solvers, and the $Tanh$ scheme described in section \ref{sec:tanh}. The results are compared to the solution given by the Roe scheme, and in the case of the void fraction, the analytical solution is shown. We can see that the high-degree dynamic interpolating polynomial $\pfp$ has an equivalent accuracy as the Roe scheme. The high-degree fixed interpolating polynomial $\Pds$ shows good accuracy. This was to be expected in this test-case as the vapor and liquid velocities are large. Thus, the intermediate eigenvalues whose orders of magnitude are the fluid velocities are in a range where $\Pds$ approximates accurately the absolute value function, yielding an accurate result. The $Tanh$ scheme also has an equivalent accuracy as the Roe scheme but, due to the computation of the hyperbolic tangent of a matrix, the computational cost is high: the computation cost is 145 times larger for the $Tanh$ than for the $\pfp$ scheme.

\subsection{Boiling channel}

The test-case consists in a one-dimensional vertical channel of length $L_h=3.65$m with upward flowing water \cite{bartolomei1982experimental, zhou2001modeling}. A uniform heat flux is imposed along the wall of the channel and causes the appearance of vapor. Two cases are considered: at the entrance, the water can be either saturated in vapor or be subcooled (i.e. be colder than the saturation temperature where vapor starts to form). In the first case, vapor creation starts at the inlet. In the second case, vapor creation starts further in the channel, when the saturation is reached, for $y=y_{boil}$. This point can be estimated analytically and is $y_{boil}=1.21\text m$, with the data used in the present test-case. This test-case checks the ability of the scheme to deal with a large range of volume fractions and to capture the onset of boiling $y_{boil}$ in the subcooled case correctly. The physics includes stiff source terms and couples hydraulics with wall heating.

The model is the \setfm presented in section \ref{sec:setfm}. Physical sources include drag force, wall friction, mass and heat transfer, and gravity. The detailed expressions of these source terms are given in appendix \ref{append:BC}. The computation is implicit. We used a one-dimensional mesh with 150 cells. We show the results at $t=5s$. At the inlet, the following values are fixed: $u_v =  0.7802$ m/s, $u_\ell = 0.7802$ m/s, $h_v = 2.784e6$ kJ/kg. The liquid enthalpy is $h_\ell = 1262$ kJ/kg in the saturated case and $h_\ell = 1029$ kJ/kg in the subcooled case (it corresponds to a subcooling of $\Delta T = 45^{\text o}$C, i.e. the temperature is lower by $45^{\text o}$C to the saturation temperature at which vapor starts to appear).
 The inlet fluid is supposed to be pure water. Thus, the initial and inlet vapor volume fractions $\avi$ will be set as small as possible. At the outlet, the pressure is fixed: $p_{outlet} = 68.73 \, 10^5 \text{ Pa}$.

\subsubsection{Boiling channel: saturated case}
\label{sec:polyn_bcs}

In the saturated boiling channel test-case, the heating sparks the creation of vapor from the inlet of the channel. The volume fraction range goes from zero to $0.95$. In practice, we will try to set the initial and inlet vapor volume fraction $\avi$ as close to zero as possible.
This case is a good demonstration of the relevance of polynomial schemes for phase appearance or disappearance. Indeed, the standard Roe scheme breaks down when vapor volume fractions are smaller than $\avi = 10^{-3}$. The polynomial schemes $\Pds$ and $\pfp$, and the $Tanh$ scheme, have no problem whatsoever even for volume fractions as small as $\avi = 10^{-8}$. 

\begin{figure}
  \centering
  \subfigure[Void fraction - $\avi = 10^{-3}$ \label{fig:bcs_3}]{\includegraphics[width=0.48\textwidth,trim=30 50 0 80,clip]{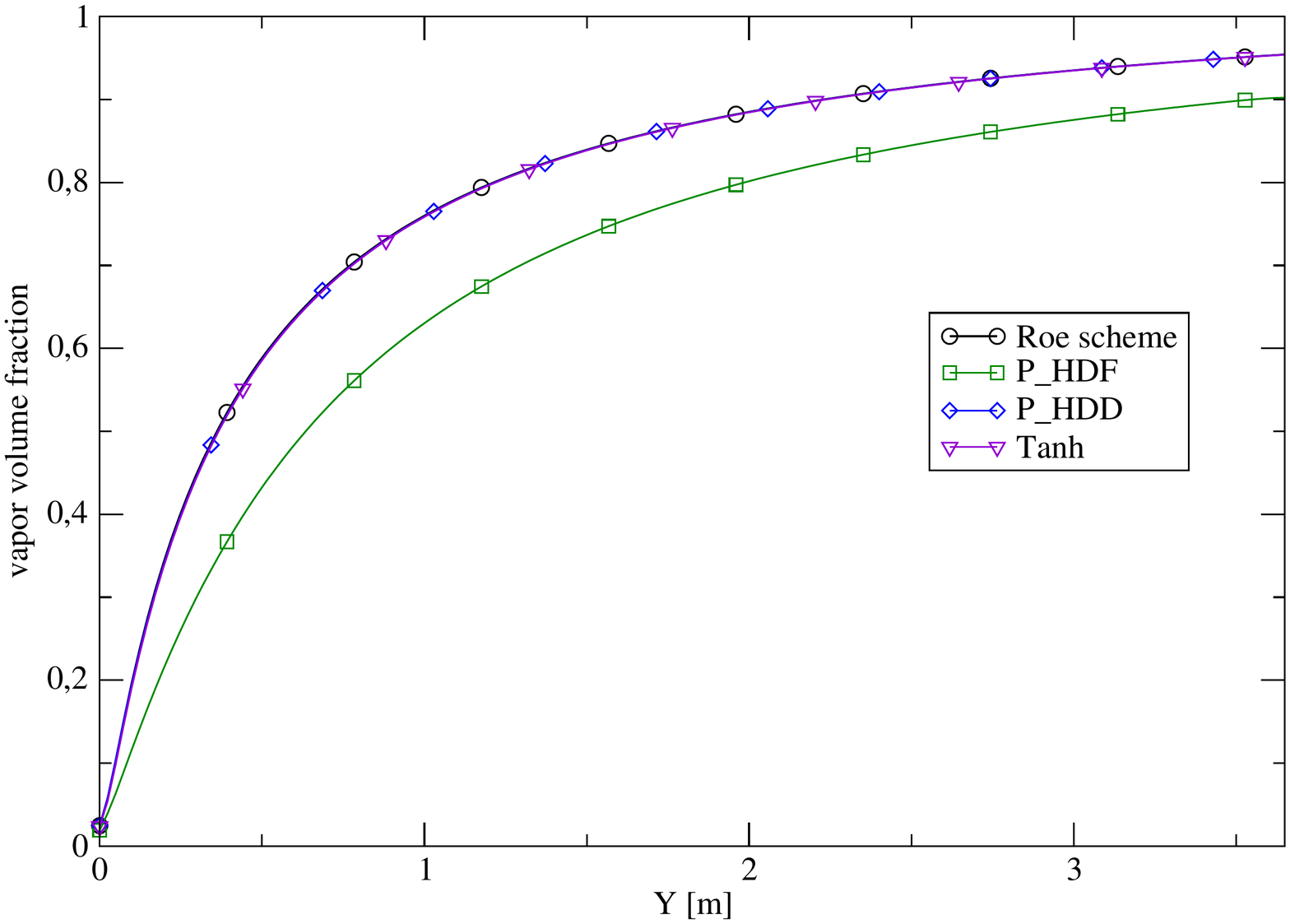}}
  \subfigure[Void fraction - $\avi = 10^{-8}$ \label{fig:bcs_8}]{\includegraphics[width=0.48\textwidth,trim=30 50 0 80,clip]{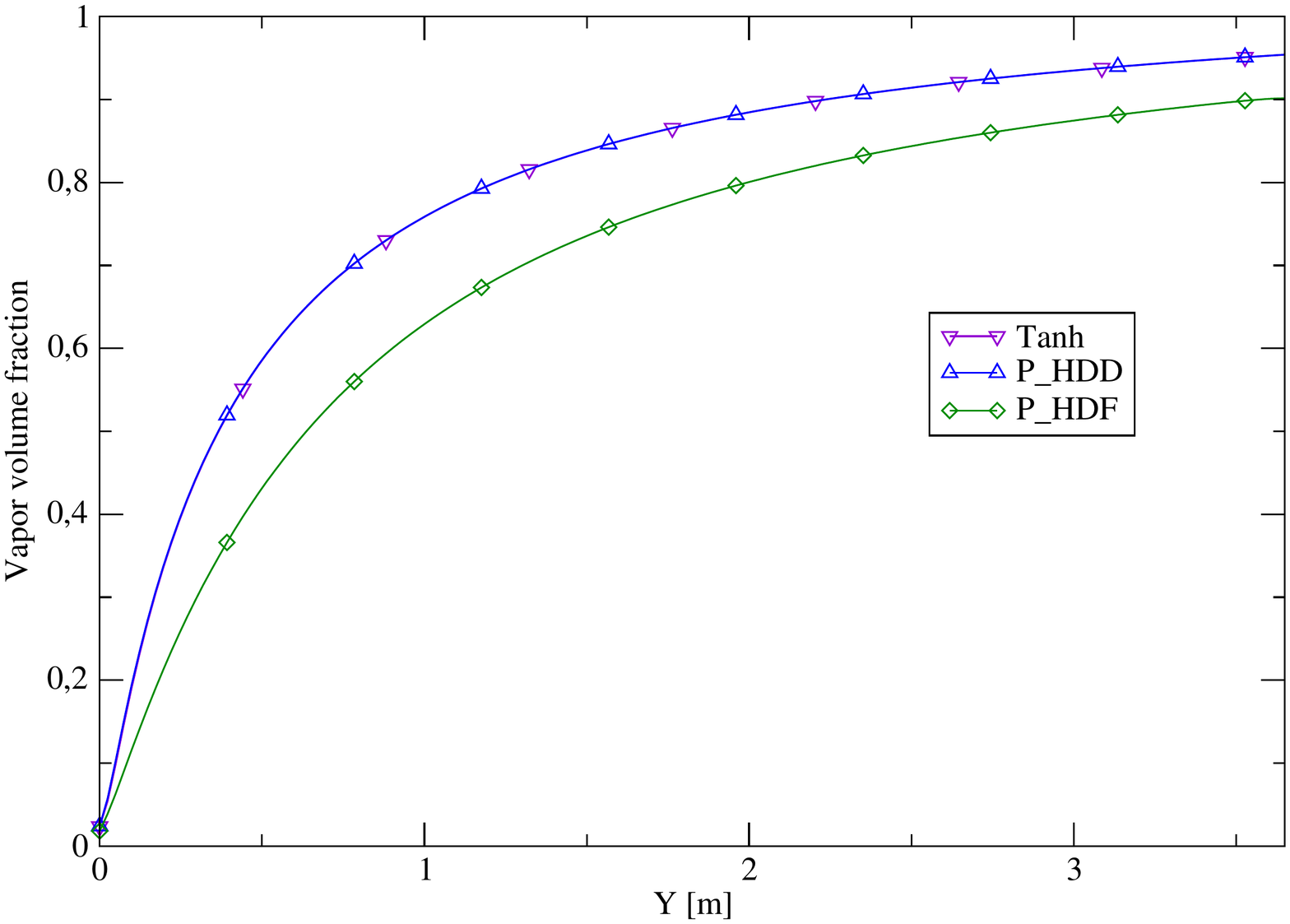}}
  \caption{Boiling channel in the saturated case with 150 cells. Void fraction for $\alpha_v = 10^{-3}$ (a) and for $\alpha_v = 10^{-8}$ (b) at time $t=5$ s as a function of space. Results by the $\Pds$ (green squares) and $\pfp$ (blue diamonds) polynomial schemes, and the $Tanh$ (violet triangles) method. The Roe scheme (black circles) is depicted on fig. (a) but breaks down in case (b). }
  \label{fig:bcs_polyn08}
\end{figure}

The profile of the vapor volume fraction is shown on fig. \ref{fig:bcs_polyn08}. As the flow is saturated, the vapor volume fraction starts increasing at the inlet of the channel. We can see on fig. \ref{fig:bcs_3} (a) for $\avi = 10^{-3}$ that the $\pfp$ and $Tanh$ methods have the same accuracy as the Roe scheme, while $\Pds$ is significantly more diffusive. This is due to the large discrepancy between the extremal and intermediate eigenvalues, which are of the order of magnitude given by eq. (\ref{eq:order_eigen}). In this case, the intermediate eigenvalues are not approximated very accurately by the polynomial $\Pds$ and the result is diffusive.
We show on fig. \ref{fig:bcs_8} (b) the void fraction profile obtained by the $\Pds$, $\pfp$ and $Tanh$ schemes for $\avi = 10^{-8}$. The positivity treatment is not needed on this case.

\subsubsection{Boiling channel : subcooled case}

\begin{figure}
  \centering
  \subfigure[Void fraction - $\avi = 10^{-2}$\label{fig:bcss_2}]{\includegraphics[width=0.48\textwidth,trim=30 50 0 80,clip]{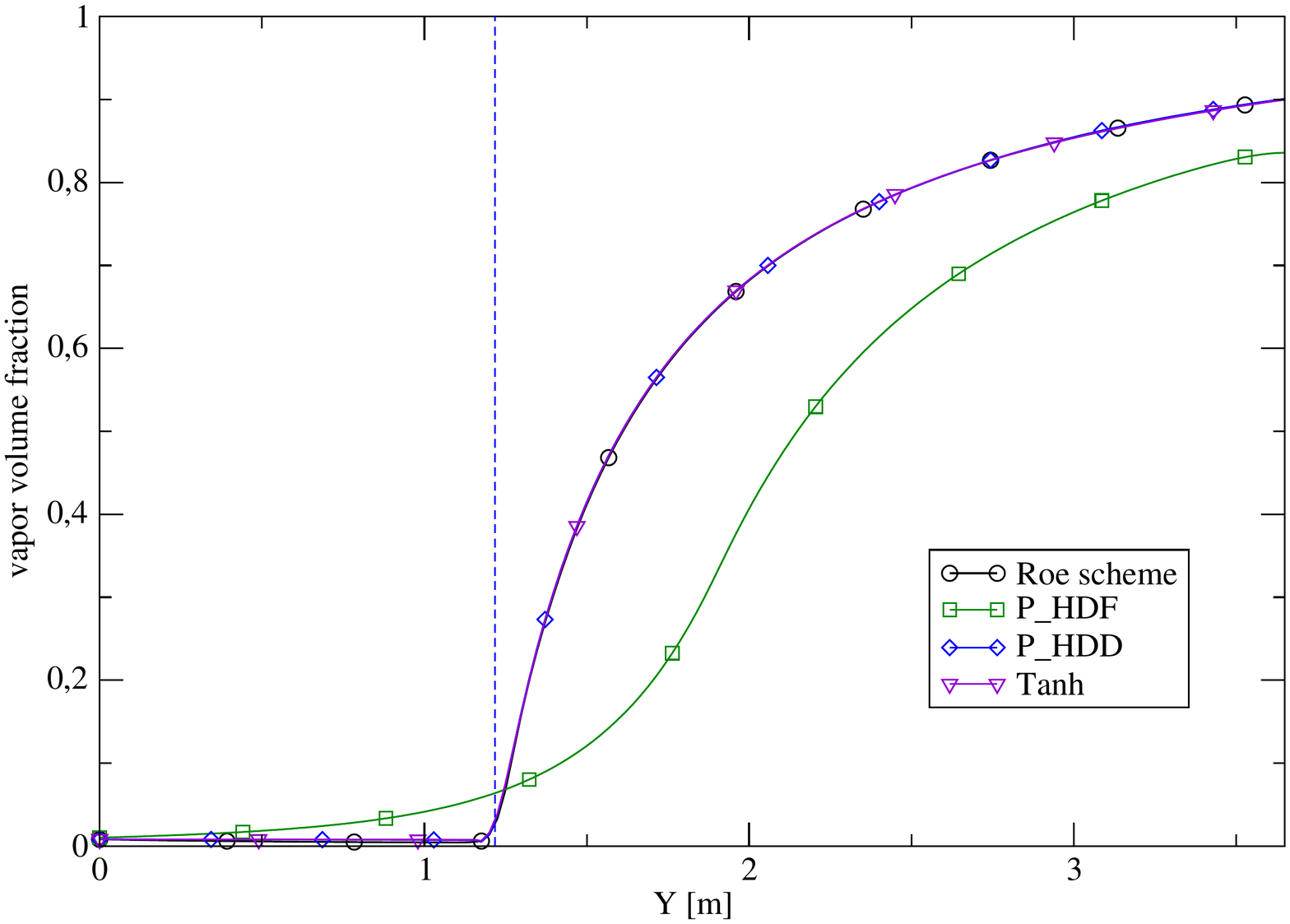}}
  \subfigure[Void fraction - $\avi = 10^{-7}$\label{fig:bcss_7}]{\includegraphics[width=0.48\textwidth,trim=30 50 0 80,clip]{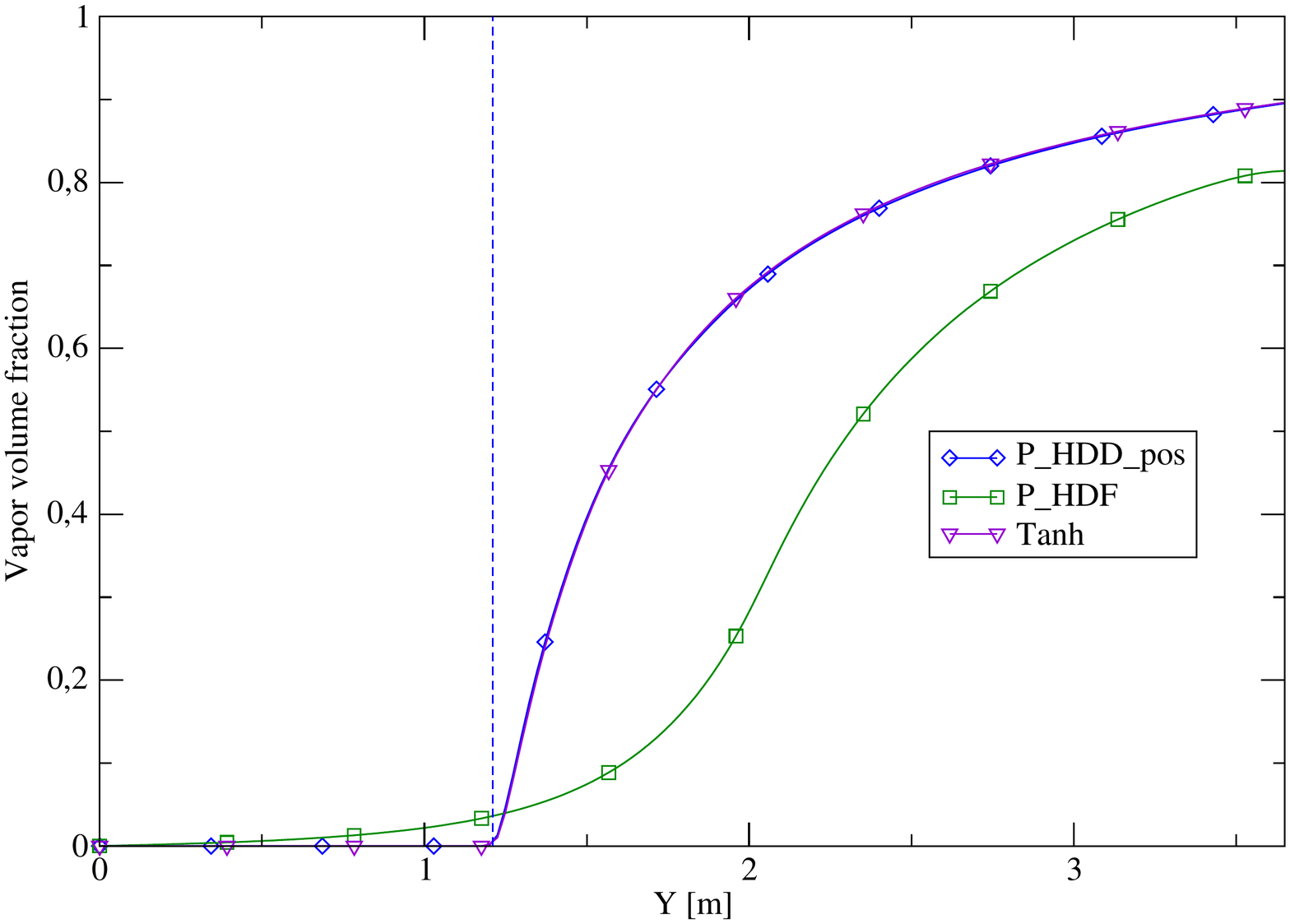}}
  \caption{Boiling channel in the subcooled case with 150 cells. Void fraction for $\alpha_v = 10^{-2}$ (a) and for $\alpha_v = 10^{-7}$ (b) at time $t=5$ s as a function of space. Results by the $\Pds$ (green squares) and $\pfp$ (blue diamonds) polynomial schemes, and the $Tanh$ (violet triangles) method. The Roe scheme (black circles) is depicted on fig. (a) but breaks down in case (b). The vertical dashed line indicates the boiling point $y_{boil}=1.21\text m$.}
  \label{fig:bcss_pfppos08}
\end{figure}

In the subcooled boiling channel case, the vapor starts being created when the saturation is reached, at the boiling point $y_{boil}=1.21\text m$. This case is more difficult than the saturated case because fluctuations are created at the boiling point and positivity problems often occur at this position. On this case, the Roe scheme presents problems for void fractions smaller than $\avi=10^{-2}$. The profile of the vapor volume fraction for an inlet and initial vapor volume fraction of $\avi=10^{-2}$ is displayed on fig. \ref{fig:bcss_2} for the Roe scheme and the $\Pds$ and $\pfp$ polynomial schemes. We can see that the vapor starts increasing when the boiling point is reached. The $\pfp$ scheme is as accurate as the Roe scheme and captures the correct boiling point, while the $\Pds$ scheme is more diffusive as in the saturated boiling channel case and the boiling point obtained by the $\Pds$ scheme is inaccurate. 

Without the positivity treatment, the polynomial scheme $\pfp$ also meets some positivity problems for vapor volume fractions smaller than $\av = 10^{-3}$. The $\Pds$ polynomial scheme is more robust due to its diffusivity and allows to compute the case for $\avi=10^{-7}$, but the result is not accurate enough to be satisfactory (fig. \ref{fig:bcss_7}). To obtain an accurate result even for small void fractions, we therefore use the polynomial scheme $\pfp$ additionned with the positivity treatment developed in section \ref{sec:pos}, called $\pfppos$ : when a positivity problem appears, the step is computed with more numerical diffusion locally where the problem is encountered. We can verify on this subcooled boiling channel test-case that positivity problems are solved whenever they appear, allowing to compute the test-case with very small vapor volume fractions while keeping the result accurate.  The void fraction profile is displayed on fig. \ref{fig:bcss_7} for the $\pfppos$ method with $\alpha_v = 10^{-7}$. As the diffusion is increased very locally, i.e. only on the faces whose neighbouring cells present a lack of positivity, the result remains very accurate and the boiling point is correctly captured. 
 The $Tanh$ method gives a very good result in terms of stability as it is able to compute the test-case with $\avi=10^{-7}$ without the positivity treatment and with a very good accuracy. However, the computational cost is very high due to the computations of the hyperbolic tangent of matrices: the computational time is multiplied by 85 on this case compared to the $\pfp$ polynomial scheme. The stability properties of the $Tanh$ scheme are thus very promising but at the present time it cannot be used in practice due to its high computational cost.

In tab. \ref{tab:sbc}, we provide some statistics on the method $\pfppos$ for different initial and inlet vapor volume fractions, and for different time-steps of the implicit computation. The subcooled boiling channel test-case has been run with CFL=10 and CFL=30, where a CFL of $1$ corresponds to the stability condition for an explicit scheme. With $a_{\max}$ the maximum signal speed, we have:
\begin{equation}
 \Delta t = CFL \frac{\Delta x}{|a_{\max}|}. 
\label{eq:cfl}
\end{equation} 
First, we can see on tab. \ref{tab:sbc} that the total number of time-steps where positivity problems or difficulties of computation of the pressure have appeared remains small: less than $0.2$\% with CFL=$10$, and less than $1.8$\% with CFL=$30$. The influence of the time-step $\Delta t$ on the occurrences of positivity problems is very clear, as there is a significant increase of problematic time-steps for a CFL of $30$ compared to a CFL of $10$. At each problematic time-step, the method $\pfppos$ has to iterate until the right diffusion is found so that the scheme is positive. We observe that the average number of iterations is close to $1$. This means that in most of the cases, the positivity problem is solved at the first iteration, i.e. with $D=10$. More rarely, it takes more than one iteration to obtain the positivity. If increasing the diffusion does not solve the positivity problem on a cell, the time-step $\Delta t$ has to be reduced (by dividing it by $10$). We can see that the time-step seldom has to be reduced.

\begin{table}[]
\centering
 \begin{tabular}{|l |c |c |c |c| c |c|} \hline
   Vapor volume fraction $\av$ &  $10^{-4}$ &$10^{-5}$& $10^{-6}$& $10^{-7}$& $10^{-8}$  \\ \hline \hline
\textbf{CFL=10}& & & & &    \\  \hline
Number of problematic time-steps  &  	0.18&	0.07&	0.03&	0.07&	0.09\\  	
 (in \% of the total number of time-steps) & 		&	& &	&		\\ \hline
Average number of iterations& 1.05	&1.00	&1.43&	1.05&	1.04\\	
per problematic time-step	& 		&	& &	 & 	\\ \hline
Number of time-steps where the time-step $\Delta t$&				1&	0&	1&	0&	0	\\
had to be divided by $10$ to obtain positivity	& 		&	&	&	&	\\ \hline \hline
\textbf{CFL=30}& & & & &    \\   \hline
Number of problematic time-steps  &  0.22&	0.22&	0.31&	1.78&	1.05\\  	
 (in \% of the total number of time-steps) &  &	&	&	&		\\ \hline			
\end{tabular} 
\caption{Statistics on the subcooled boiling channel test-case over 5s of computation ($\approx 50 000$ iterations)  for the $\pfppos$ scheme}
\label{tab:sbc}
\end{table}

\subsection{Tee junction}

The two-dimensional tee-junction test-case shows a dynamic separation between the liquid and the vapor phase, thus creating accumulation of vapor in some spots and disappearance in others. It consists in a two-dimensional horizontal pipe $T_1$ of length $0.877$ $m$ and diameter $0.055$ $m$, connected to an other horizontal pipe $T_2$ of diameter $0.055$ $m$ in $x=0.197$ $m$ and whose length from the junction is $0.7196$ $m$. A mixture of water and steam enters the first pipe $T_1$ at $x=(0.0,0.0)$. Due to the difference of density  and thus inertia between vapor and liquid, most of the liquid continues in the first pipe after the junction while a big part of the vapor is deported in the second pipe at the junction. Vapor thus accumulates at the junction. The phenomenon is only dynamic as no phase change occurs in this test-case. 
This test-cases allows to test the ability of the scheme to deal with a large volume fraction range.

\begin{figure}
  \centering
  \subfigure[Coarse mesh \label{fig:tee_c_mesh}]{\includegraphics[width=0.48\textwidth]{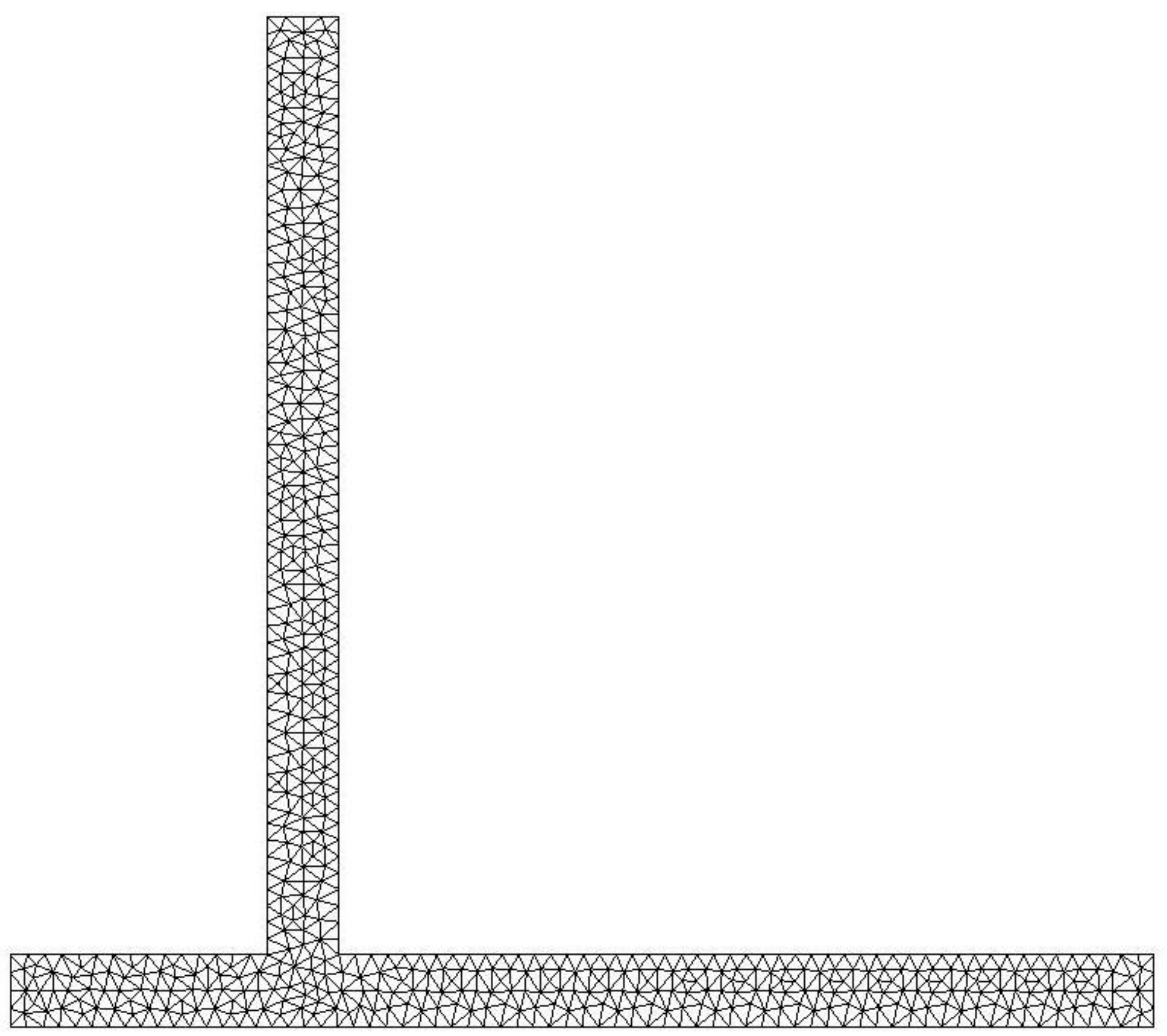}}
  \subfigure[$\pfppos$ - Void fraction]{\includegraphics[width=0.5\textwidth]{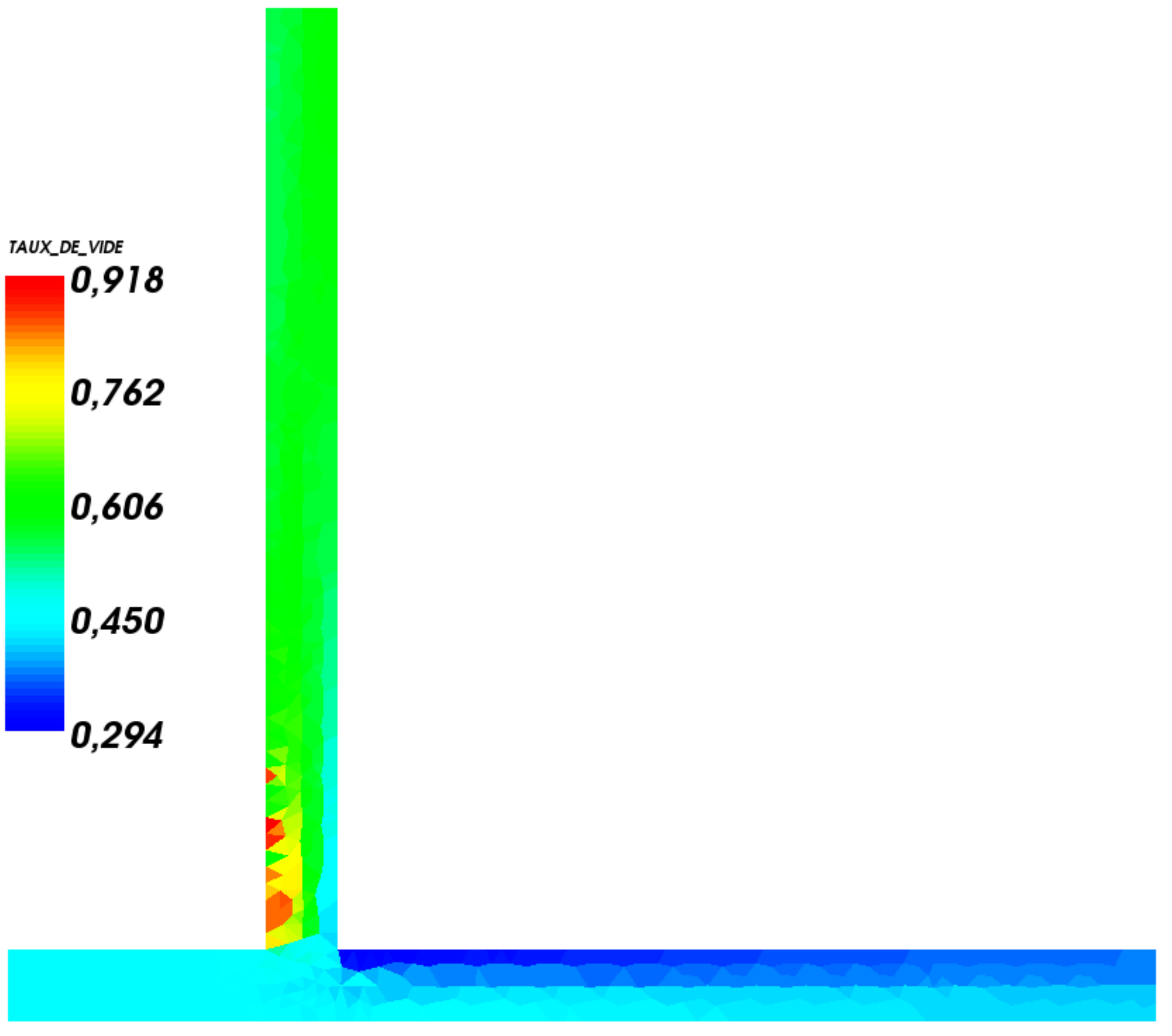}}
  \caption{Tee junction computed by the $\pfp$ scheme on the coarse mesh. Left: mesh used for the computation. Right: vapor volume fraction as a function of space (color coded). }
  \label{fig:tee_coarse}
\end{figure}

\begin{figure}
  \centering
  \subfigure[Refined mesh \label{fig:tee_r_mesh}]{\includegraphics[width=0.48\textwidth]{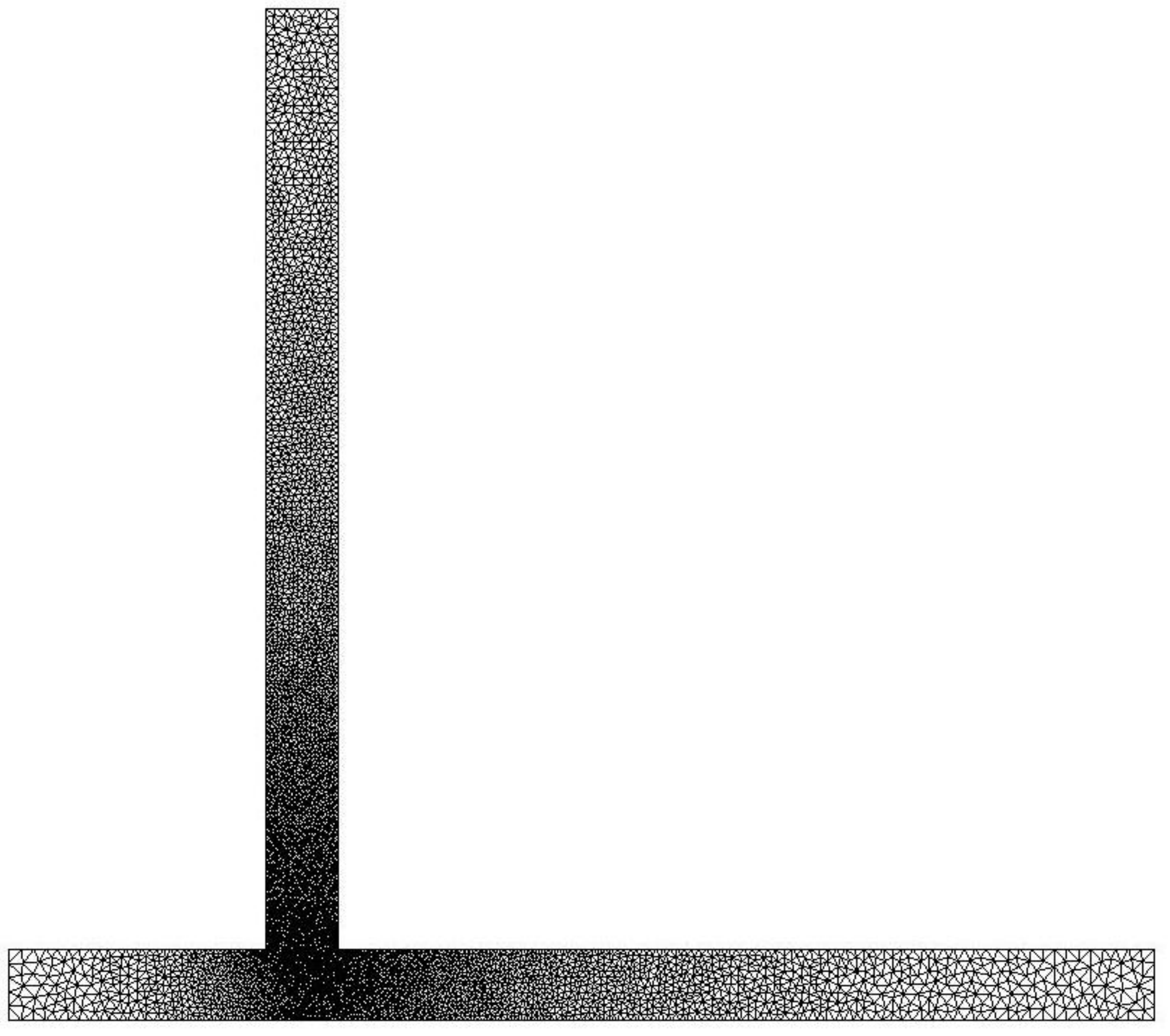}}
  \subfigure[$\pfppos$ - Void fraction]{\includegraphics[width=0.5\textwidth]{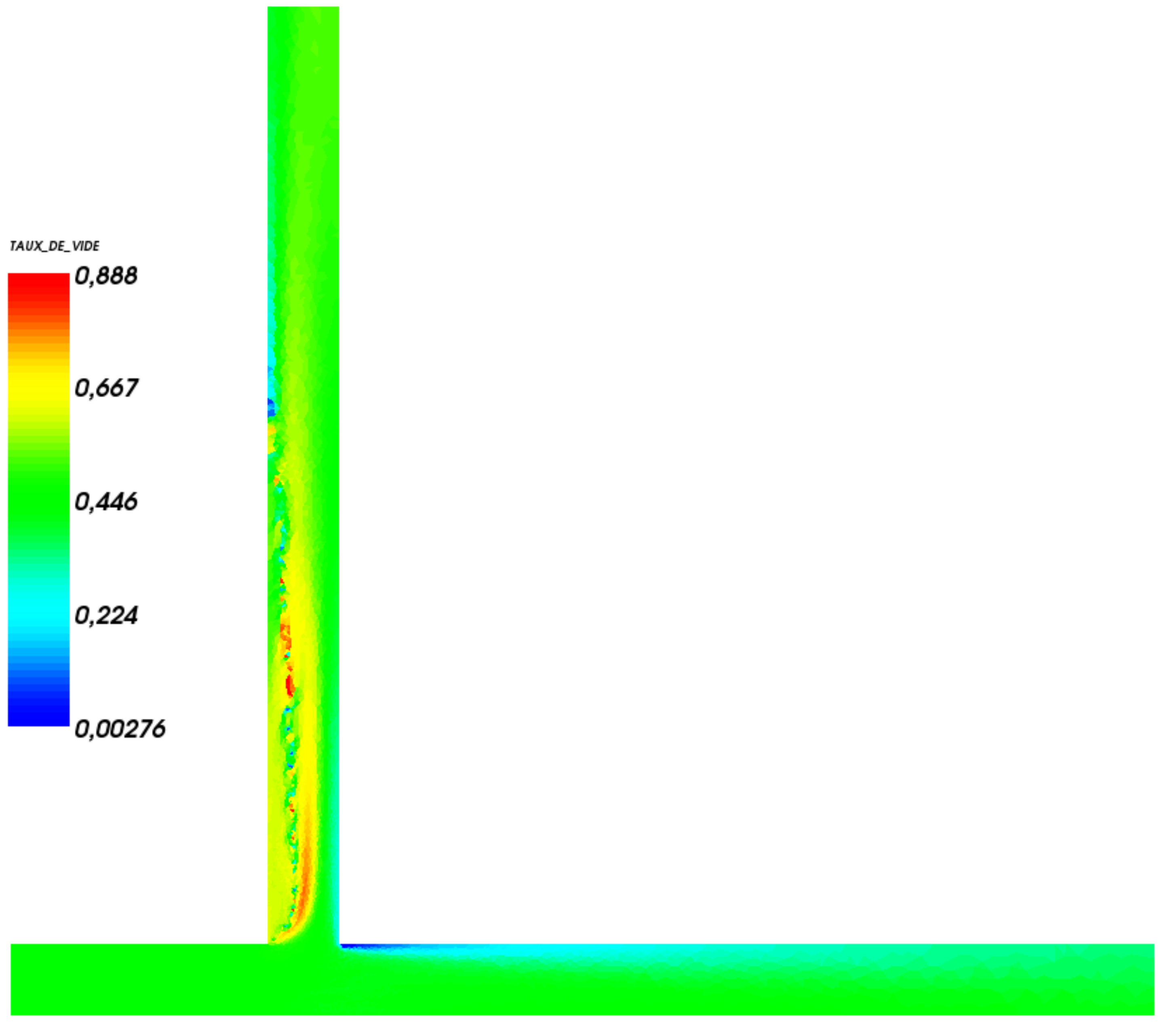}}
  \caption{Tee junction computed by the $\pfppos$ scheme on the refined mesh.  Left: mesh used for the computation. Right: vapor volume fraction as a function of space (color coded).}
  \label{fig:tee_ref}
\end{figure}

The model is the two-fluid two-phase flow model presented in section \ref{sec:setfm}. The source terms included in the case are detailed in appendix \ref{append:TJ}. At the inlet, the following values are fixed: $u_v =  (1.0, 0.0)$ m/s, $u_\ell = (1.0, 0.0)$ m/s, $h_v = 2650$ kJ/kg, $h_\ell = 1607$ kJ/kg and $\alpha_v=0.45$. 
The pressure is fixed at the outlet. At the outlet of the horizontal pipe, $p^{outlet}_1 = 150 \, 10^5 \text{ Pa}$, and at the outlet of the vertical pipe, $p^{outlet}_2 = 149.998 \, 10^5 \text{ Pa}$. A wall slip boundary condition is prescribed on the walls. The computation is implicit. A coarse mesh with 1149 cells (fig. \ref{fig:tee_c_mesh}) and a refined mesh with 11315 cells (fig. \ref{fig:tee_r_mesh}) have been used.

This case cannot be run with the Roe scheme as positivity problems are met with the coarse and the refined meshes. A first improvement is brought by the use of polynomial schemes as the $\pfp$ scheme is able to compute the case on the coarse mesh without any problem. The result for the void fraction is shown on fig. \ref{fig:tee_coarse}. The result obtained by the $\Pds$ scheme is too diffusive to be of interest. However, the $\pfp$ scheme alone is not able to compute the case on the refined mesh, due to positivity problems. Therefore we use the positivity treatment developed in section \ref{sec:pos}. All the positivity problems are overcome by this method and we are able to show the result of the computation with $\pfppos$ on fig. \ref{fig:tee_ref}, for the vapor volume fraction.

In tab. \ref{tab:teeref}, we provide some statistics on the $\pfppos$ method for the tee-junction test-case on a refined mesh. During the computation, the CFL (eq. (\ref{eq:cfl})) increases linearly in a lap time of $10$ s from CFL=$50$ to CFL=$690$. Tab. \ref{tab:teeref} shows that the number of time-steps where positivity problems appear remains very small. 
The average number of iterations of the algorithm remains below two iterations. It means that most of the time the positivity problems are solved in one iteration of the $\pfppos$ method, i.e. with a diffusion $D=10$. Only two time-steps have required a diminution of the time step $\Dt$ in order to obtain the positivity.

\begin{table}[h!]
\centering
 \begin{tabular}{|l| c| c| }
\hline  
Number of problematic time-steps  &  0.036 	 \\  	
 (in \% of the total number of time-steps) & 		\\ \hline 
 Average number of iterations&     1.67          	\\	
 per problematic time-step	&     \\  \hline 
Number of time-steps where the time-step $\Delta t$ &     2            	\\	
had to be divided by $10$ to obtain positivity	&    \\ \hline 
\end{tabular} 
\caption{Statistics on the tee-junction test-case with a refined mesh and the $\pfppos$ scheme, between 0s and 5s (85900 iterations).
}
\label{tab:teeref}
\end{table}

\section{Conclusion}
\label{sec:conclu}

In this paper, we have considered numerical schemes for multi-phase flow models when one of the phases appears or disappears. The causes of the difficulties that standard methods face in this situation have been identified. They are: (i) the loss of hyperbolicity of the model when a phase appears or disappears ; (ii) the lack of positivity of the scheme. Polynomial schemes have been developed to avoid the use of the eigenvector decomposition of the Roe matrix and tackle problem (i). A specific positivity treatment has been applied to the polynomial solver to treat problem (ii). The resulting method is very robust: large ranges of void fraction can now be computed with high accuracy. The method has proved effective and accurate on test problems on which standard methods fail. An alternate method, based on the hyperbolic tangent function, has also been proposed. It is as accurate as the polynomial solver and has shown very  good positivity properties without requiring any positivity treatment. However, it is computationally too intensive. Future work will be concerned with improving the computational cost of the hyperbolic tangent method, and combining the polynomial method with the all-speed methodology proposed in \cite{CDK, DT}. The latter will allow to treat situations where some parts of the flow are in the small Mach-number regime.



\appendix

\section{Appendix}
\label{sec_appendix}

\subsection{Eigenvalues of the two-fluid model}
\label{append:eigenvalues}

We investigate the structure of the eigenvalues of the two-fluid system (including the energy equations). We recall the method employed in \cite{TK96, KTC}. In the finite volume framework, the system can be written in the quasilinear form:
\begin{equation*}
  \frac{\partial \V}{\partial t} + \A_{\mathbf n}(\V) \frac{\partial \V}{\partial \mathbf n} = 0.
\end{equation*}
where $\mathbf n$ is the normal vector on the considered face. We look for the roots of a polynomial $P(\lambda)$, the characteristic polynomial of the $\A_{\mathbf n}$ matrix, of degree $2(2+d)$, $d$ being the space dimension.
A straightforward computation leads to the following polynomial
$$ P(\lambda) = (\lambda - u_{v n})^d (\lambda - u_{l n})^d P_4(\lambda) $$ 
where $u_{kn}$ is the the projection on the normal vector of the velocity of phase $k$, and $P_4$ is a polynomial of degree $4$.
It follows immediately that $u_{v n}$ and $u_{l n}$ are some of the eigenvalues of the system of multiplicity $d$.

For the other eigenvalues, we look for an approximation of the roots of $P_4$ and use a perturbation method by introducing the small ratio
\begin{equation}
 \xi=\frac{u_{r n}}{a_m},
\label{eq:xi} 
\end{equation}
where $u_{r n}$ is the projection of the relative velocity on the normal vector and $a_m$ is the 'characteristic' speed of sound, in the two-phase mixture, given by
\begin{eqnarray*} 
  a_m & = & \left( \frac{\rho _m (\alpha _v \rho _\ell + \alpha _\ell \rho 
    _v)}{\rho _v \rho _\ell} \right)^{1/2}, \qquad c_m 
  =  \left( \frac{\alpha_\ell\rho_v+\alpha_v\rho_\ell}{\alpha_\ell\rho_v c_\ell^{-2} 
    +\alpha_v\rho_\ell c_v^{-2} }  \right)^{1/2}
\end{eqnarray*} 
with $c_m$ the mixture sound velocity : $  c_m^2 = \frac{\rv \rl }{\rho_m } \gamma^2$
and
\begin{equation*}
  \gamma^2=\frac{c_v^2 c_\ell^2}{\av\rl c_\ell^2+\al\rv c_g^2}, \qquad \frac{1}{c_k^2} =\left(\frac{\partial\rk}{\partial p}\right)_{s_k},  
\end{equation*}  
and $s_k$ is the entropy. The first order approximation of the two-fluid system eigenvalues is  
\begin{equation} 
  \left \{ 
  \begin{array}{ll} 
    {\displaystyle 
      \frac{\alpha_v \rho_\ell u_{v n} + 
        \alpha_\ell \rho_v u_{l n}}{\alpha_\ell \rho_v + 
        \alpha_v \rho_\ell} - a_m + O(\xi^2) ,
    }  \\ 
    {\displaystyle 
      \frac{\alpha_v \rho_\ell u_{vn} + 
        \alpha_\ell \rho_v u_{ln}}{\alpha_\ell \rho_v + 
        \alpha_v \rho_\ell} + a_m + O(\xi^2) ,
    }\\ 
    {\displaystyle 
      \frac{\alpha_\ell \rho_v u_{vn} + \alpha_v \rho_\ell 
        u_{ln}}{\alpha_\ell \rho_v + \alpha_v \rho_\ell} 
      - 
      \sqrt{ \frac{1}{\alpha_\ell \rho_v + \alpha_v \rho_\ell}(\Delta P^i - \frac{u_{rn}^2 \alpha_v \rho_v \alpha_\ell \rho_\ell}{\alpha_\ell \rho_v + \alpha_v \rho_\ell})} 
      + O(\xi^2) ,
    }\\ 
    {\displaystyle 
      \frac{\alpha_\ell \rho_v u_{vn} + \alpha_v \rho_\ell 
        u_{ln}}{\alpha_\ell \rho_v + \alpha_v \rho_\ell} 
      + 
      \sqrt{ \frac{1}{\alpha_\ell \rho_v + \alpha_v \rho_\ell}(\Delta P^i - \frac{u_{rn}^2\alpha_v \rho_v \alpha_\ell \rho_\ell}{\alpha_\ell \rho_v + \alpha_v \rho_\ell})} 
      + O(\xi^2) ,
    }  \\ 
  \end{array} 
  \right.
  \label{eq:vp}
\end{equation} 
with $\Delta P^i$ the interfacial pressure default. The approximate formula of the eigenvalues associated with the void waves leads to the hyperbolicity condition  
\begin{equation*} 
  \Delta P^i \ge \frac{({\mathbf u}_r \cdot {\mathbf n})^2\alpha_v \rho_v \alpha_\ell \rho_\ell}{\alpha_\ell \rho_v + \alpha_v \rho_\ell} ,
\end{equation*}
which corresponds to Bestion's model for the interfacial pressure term \cite{bestion1990physical}. Expressions (\ref{eq:vp}) can be written in the following form
\begin{equation*} 
  \left \{ 
  \begin{array}{ll} 
    {\displaystyle 
      ({\mathbf u}_v - \frac{\kappa \alpha_\ell}
      {\alpha_v +\alpha_\ell\kappa}  {\mathbf u}_r) \cdot {\mathbf n}
      - a_m + O(\xi^2) ,
    }  \\ 
    {\displaystyle 
      ({\mathbf u}_v - \frac{\kappa \alpha_\ell}
      {\alpha_v +\alpha_\ell\kappa}  {\mathbf u}_r) \cdot {\mathbf n}
      + a_m + O(\xi^2) ,
    }\\ 
    {\displaystyle 
      ({\mathbf u}_\ell + \frac{\kappa\alpha_\ell}
      {\alpha_v +\alpha_\ell\kappa} {\mathbf u}_r -
      \frac{\sqrt{ (\delta - 1) \alpha_v \alpha_\ell \kappa}}{\alpha_v+\alpha_\ell \kappa} {\mathbf u}_r) \cdot {\mathbf n}
      + O(\xi^2) ,
    }\\ 
    {\displaystyle 
      ({\mathbf u}_\ell + \frac{\kappa\alpha_\ell}
      {\alpha_v +\alpha_\ell\kappa} {\mathbf u}_r +
      \frac{\sqrt{ (\delta - 1) \alpha_v \alpha_\ell \kappa}}{\alpha_v+\alpha_\ell \kappa} {\mathbf u}_r) \cdot {\mathbf n}
      + O(\xi^2) ,
    }  \\ 
  \end{array} 
  \right. \label{eq:vp2}
\end{equation*}  
with $\kappa=\frac{\rho_v}{\rho_\ell}$ denoting in general a small number.

\subsection{Void fraction and pressure wave eigenvectors of the two-fluid model, and asymptotic behaviour}
\label{append:eigenvectors}

A first-order approximation in $\xi$ (given by \ref{eq:xi}) of the eigenvectors of the two-fluid model has been given in \cite{toumi96} for a perfect gas of constant $\gamma$. Let us recall the expression of the right eigenvectors $R_3$ and $R_4$ associated to the eigenvalues $\la_3$ and $\la_4$ in (\ref{eq:vp2}), and which are suspected to collapse when the void fraction $\av$ tends to zero: 
\begin{equation} 
R_{3,4} =
  \left[ 
  \begin{array}{c} 
\displaystyle  1 \\
\displaystyle  -\frac{\rho_\ell}{\rho_v} \\
\displaystyle  \la_{3,4} \\
\displaystyle  -\frac{\rho_\ell}{\rho_v}  \la_{3,4}\\
\displaystyle  \frac{1}{\gamma}(H_v - \ud u_v^2) - u_v(\ud u_v-\la_{3,4})\\
\displaystyle  -\frac{\rho_\ell}{\rho_v} (H_\ell - \frac{p}{\rho_\ell})
  \end{array} 
  \right] . 
\end{equation} 

Let us now suppose that the vapor phase disappears and the vapor volume fraction $\av$ tends to zero. In this case, we assume that the relative velocity $\uv_{rn}$ also tends to zero. 
The fast eigenvalues $\la_1$ and $\la_2$ are now equal to $u_{n} \pm a_m$. They remain distinct and the eigenvectors associated to these eigenvalues do not collapse.
As for the intermediate eigenvalues, the void eigenvalues $\la_3$ and $\la_4$, the form of which are recalled below: 
$$\la_{3,4}= ({\mathbf u}_\ell + \frac{\kappa\alpha_\ell}
      {\alpha_v +\alpha_\ell\kappa} {\mathbf u}_r \pm
      \frac{\sqrt{ (\delta - 1) \alpha_v \alpha_\ell \kappa}}{\alpha_v+\alpha_\ell \kappa} {\mathbf u}_r) \cdot {\mathbf n}
      + O(\xi^2) , $$  
tend to $u_{n}$.

One can also check that the eigenvectors $R_3$ and $R_4$ have the following form $R= R^0 + \delta R+ O(\xi^2)$, namely:
\begin{equation} 
R=
  \left[ 
  \begin{array}{c} 
\displaystyle  1 \\
\displaystyle  -\frac{\rho_\ell}{\rho_v} \\
\displaystyle  u_{n} \\
\displaystyle  -\frac{\rho_\ell}{\rho_v} u_{n} \\
\displaystyle  \frac{1}{\gamma}(H_v - \ud u^2) +\ud u^2\\
\displaystyle  -\frac{\rho_\ell}{\rho_v} (H_\ell - \frac{p}{\rho_\ell})
  \end{array} 
  \right] \pm
  \left[ 
  \begin{array}{c} 
\displaystyle  0 \\
\displaystyle  0 \\
\displaystyle  u_{rn}\sqrt{\av} \beta \\
\displaystyle  -\frac{\rho_\ell}{\rho_v} u_{rn}\sqrt{\av} \beta\\
\displaystyle   \uv \cdot \uv_r \sqrt{\av} \beta\\
\displaystyle  0
  \end{array} 
  \right] 
+ O(\xi^2) , 
\end{equation} 
with $\beta =  \frac{\sqrt{ (\delta - 1) \alpha_\ell \kappa}}{\alpha_v+\alpha_\ell \kappa}$ and $\beta\to \sqrt{\frac{ \delta - 1 }{\kappa}}$ when $\av\to0$.
When $\av$ tends to zero, $\uv_r$ tends to zero and so does $\xi$. Therefore, $\delta R \sim \alpha^\ud u_r$ also tends to zero and $R_3$ and $R_4$ collapse.

\subsection{Test-cases}

\subsubsection{Boiling channel}
\label{append:BC}

 The model used in the boiling channel test-case is the two fluid two phase flow model presented in section \ref{sec:setfm}. Here are the modeling terms included in the case. We assume that while $h_\ell<h_\ell^{sat}$, the liquid saturation enthalpy, the heat flux is only implied in the heating of the liquid (heat transfer). When $h_\ell>h_\ell^{sat}$, the heat flux becomes implied in the evaporation only and therefore results in mass transfer.
The mass transfer also implies a transfer of momentum and energy.
All numerical values are indicated below. 

\begin{enumerate}
\item The interfacial pressure term is the Bestion's modeling term (\ref{term:bestion}) with $\delta_0 = 1.1$ and $\kappa=10^{-4}$.
\item Interfacial velocities and enthalpies: 
$$ \mathbf u^i =  \mathbf u_\ell, $$
$$h_v^i = h_v^{sat},\qquad h_\ell^i = h_\ell^{sat}. $$
\item Wall heat transfer concentrations:
\begin{align*}
Q^w_\ell & =   q \qquad \text{if} \qquad h_\ell<h_\ell^{sat},  \\
         & = 0  \qquad \text{otherwise}. \\
Q^w_v & =  0. 
\end{align*}
\item Mass transfer:
\begin{align*}
\Gamma &= 0 \qquad \text{if} \qquad h_\ell<h_\ell^{sat}, \\
       &= \frac{q}{L} \qquad \text{otherwise}.
\end{align*}
\item Drag force: 
$$ \mathbf  F^{iD}_v = -F^{iD}_\ell = -\frac{1}{8} C_D a_i\rho_m | \mathbf u_r|  \mathbf  u_r  . $$ 
\item Wall friction:
$$ \mathbf  F_k^w = \frac{f}{D_h}\frac{\alpha_k \rho_k| \mathbf  u_k| \mathbf  u_k}{2} . $$
\item Gravity:
$$ \mathbf  f_{ext} =  \mathbf  g. $$
\end{enumerate}
Numerical data and auxiliary relations are given in tables \ref{tab_bc} and \ref{tab_rel_bc}.

\noindent 
\begin{table}
\begin{center}
\renewcommand{\arraystretch}{2.5}
\begin{tabular}{|l|l|l|}
\hline
$D_h=0.628$ m & $L_h =3.65$ m & $N_{PCH}=10$ \\
\hline
$u_0 = 0.7802$ m/s &  $\displaystyle a_i= \frac{ 3 \av }{ r_i} $ with $r_i = 5.10^{-4}$ & $C_D = 0.44$ \\
\hline
$f=0.017$ & $g=-9.81$ m/$s^2$ & \\
\hline
\end{tabular}
\caption{Numerical data for the boiling channel test-case}
\label{tab_bc}
\end{center}
\end{table}

\noindent 
\begin{table}
\begin{center}
\renewcommand{\arraystretch}{3.5}
\begin{tabular}{|l|l|l|}
\hline
$\displaystyle L = h_v^{sat} - h_\ell^{sat}$ &   $\displaystyle v_{lv} = \frac{1}{\rho_v^{sat}} - \frac{1}{\rho_\ell^{sat}}$ &  $\displaystyle u_r = u_v -u_\ell$ \\
\hline
$\displaystyle \rho_m = \alpha_v \rho_v + \alpha_\ell \rho_\ell$ & $\displaystyle q =  \frac{N_{PCH} u_0 L} {L_h v_{lv}}$  & \\
\hline
\end{tabular}
\caption{Auxiliary relations for the boiling channel test-case}
\label{tab_rel_bc}
\end{center}
\end{table}

\subsubsection{Tee Junction}
\label{append:TJ}
The model used in the tee-junction test-case is the two-fluid two-phase flow model presented in section \ref{sec:setfm}. Here are the source terms included in the case:
\begin{enumerate}
\item The interfacial pressure term is the Bestion's modeling term (\ref{term:bestion}) with $\delta_0 = 1.1$ and $\kappa=10^{-4}$.
\item Interfacial velocity: 
  $$ \mathbf u^i =  \mathbf u_\ell, $$
\item Drag force: 
  $$ \mathbf  F^{iD}_v = -F^{iD}_\ell = -\frac{1}{8} C_D a_i\rho_\ell | \mathbf u_r|  \mathbf  u_r,   $$ 
  with $a_i= \frac{ 3 \av }{ r_i} $,  $r_i =  0.3165\ 10^{-3}$, and $C_D = 0.44$.
\item Wall friction:
  $$ \mathbf  F_k^w = \frac{f}{D_h}\frac{\alpha_k \rho_k| \mathbf  u_k| \mathbf  u_k}{2}, $$
  with $D_h=1$ m and $f=0.05$.
\end{enumerate}

\subsection[Coefficients of the polynomial PHDF]{Coefficients of the polynomial $\Pds$}
\label{sec:coeffp17}

The $\Pds$ polynomial is written:
$$P(x) = \sum_{k=0}^{17} a_k x^{2k}.$$
with the $a_k$ given by:

\begin{minipage}{0.5\textwidth}
  \begin{align*}
    a_0 &=    6.209633161688544e-02\\
    a_1 &=      4.516480010541272e+00\\
    a_1 &=      -3.049057345414379e+01\\
    a_2 &=      1.657256844603353e+02\\
    a_4 &=     -6.133533687894306e+02\\
    a_5 &=      1.580698142537855e+03\\
    a_6 &=     -2.879210705862515e+03\\
    a_7 &=      3.673105197391366e+03\\
    a_8 &=      -3.121407591514732e+03\\
  \end{align*}
\end{minipage} \hfill
\begin{minipage}{0.5\textwidth}
  \begin{align*}
    a_9 &=       1.512887040780976e+03\\
    a_{10} &=      -2.111058506112595e+02\\
    a_{11} &=      9.753698909265717e+01\\
    a_{12} &=     -6.475861637079317e+02\\
    a_{13} &=      8.947647548149256e+02\\
    a_{14} &=      -6.303841204016171e+02\\
    a_{15} &=       2.586951712420909e+02\\
    a_{16} &=      -5.941358894806618e+01\\
    a_{17} &=       5.960406627331660e+00\\
  \end{align*}
\end{minipage}


\end{document}